\documentclass[aip,a4paper,amsmath,amssymb,reprint]{revtex4-2}
\usepackage{amsmath,amssymb,bm,graphicx,xcolor,multirow,commath,tabularray,etoolbox}
\usepackage[colorlinks,citecolor=blue,linkcolor=red,urlcolor=blue]{hyperref}
\usepackage[caption=false]{subfig}
\usepackage[utf8]{inputenc}
\usepackage[T1]{fontenc}
\begin{document}

\title[Non-linear growth of viscously-unstable two-phase flow
fingers]{Disorder-induced non-linear growth of viscously-unstable immiscible
two-phase flow fingers in porous media}

\author{Santanu Sinha} 
\email{santanu.sinha@ntnu.no}
\affiliation{PoreLab, Department of Physics, University of Oslo, N--0316 Oslo, Norway\looseness=-1}

\author{Yves M{\'e}heust}
\email{yves.meheust@univ-rennes1.fr}
\affiliation{CNRS, G{\'e}osciences Rennes, UMR 6118, Universit{\'e} de Rennes 1, F--350042 Rennes, France\looseness=-1}

\author{Hursanay Fyhn}
\email{hursanay.fyhn@sintef.no}
\affiliation{SINTEF Energy Research, Trondheim N-7465, Norway\looseness=-1}
\affiliation{PoreLab, Department of Physics, Norwegian University of Science and Technology, N--7491 Trondheim, Norway\looseness=-1}

\author{Subhadeep Roy}
\email{subhadeep.r@hyderabad.bits-pilani.ac.in}
\affiliation{Department of Physics, Birla Institute of Technology and Science Pilani, Hyderabad Campus, Telangana 50078, India\looseness=-1}

\author{Alex Hansen}
\email{alex.hansen@ntnu.no}
\affiliation{PoreLab, Department of Physics, Norwegian University of Science and Technology, N--7491 Trondheim, Norway\looseness=-1}

\date{\today}

\begin{abstract}
    The immiscible displacement of a fluid by another one inside a porous medium
    produces different types of patterns depending on the capillary number Ca
    and viscosity ratio $M$. At high Ca, viscous fingers resulting from the
    viscous instability between fluid-fluid interfaces are believed to exhibit
    the same Laplacian growth behavior as viscously-unstable fingers observed in
    Hele-Shaw cells by Saffman and Taylor \cite{st58}, or as diffusion limited
    aggregates (DLA) \cite{p84}. I.e., the interface velocity depends linearly
    on the local gradient of the physical field that drives the growth process
    (for two-phase flow, the pressure field). However, steady-state two-phase
    flow in porous media is known to exhibit a regime for which the flow rate
    depends as a non-linear power law on the global pressure drop, due to the
    disorder in the capillary barriers at pore throats. A similar nonlinear
    growth regime was also evidenced experimentally for viscously-unstable
    drainage in two-dimensional porous media 20 years ago \cite{lmt04}. Here we
    revisit this flow regime using dynamic pore-network modeling, and explore
    the non-linearity in the growth properties. We characterize the
    previously-unstudied dependencies of the statistical finger width and
    nonlinear growth law's exponent on Ca, and discuss quantitatively, based on
    theoretical arguments, how disorder in the capillary barriers controls the
    growth process' non-linearity, and why the flow regime crosses over to
    Laplacian growth at sufficiently high Ca. In addition, the statistical
    properties of the fingering patterns are compared to those of Saffman-Taylor
    fingers, DLA growth patterns, and the results from the aforementioned
    previous experimental study. 
\end{abstract}


\maketitle

\section{Introduction}
Fingering patterns are one of many unique features of two-phase flow which are
caused by hydrodynamic instabilities between two fluids \cite{cmp59, yh86}. When
one fluid displaces another fluid inside a medium, depending on the properties
of the two fluids and the medium in which they are flowing, the displacement
front may exhibit fingering instead of a stable interface between the two
fluids, hereafter denoted the front \cite{bbo2015,md86,fms97}. A wide variety of
fingering patterns are observed for different types of multiphase flow, such as
the flow of miscible \cite{pjh65,p85,jcj11,fcj17} and immiscible
\cite{lmt04,sss22} fluids in continuum \cite{st58,cmp59} and porous medium
\cite{b98,ffh22,par21}, reactive transport flow \cite{sfz23} and the flow of
frictional fluids in granular materials \cite{sfk11,zce23}. The structural
properties of different types of fingers are controlled by the underlying
physical forces, such as the viscous, capillary, inertial and frictional forces.
Furthermore, depending on the driving condition or the geometry of the system,
the displacement front can also undergo transition from fingering patterns to
compact \cite{emj22} or foam-like structures \cite{lsh23}.

Our study here deals with non-reactive two-phase flow in non-deformable medium,
where the two fluids are immiscible and separated by interfaces associated with
a surface tension. Inside a continuum medium, for example a Hele-Shaw cell,
which consists of two parallel plates separated by a small gap, the displaced
fluid having a higher viscosity than the displacing fluid leads to viscous
instability of the front and the development of fingers with a constant width,
except in the vicinity of the finger tip. The smooth rounded shape of these
fingers in a two-dimensional $zx$ plane can be described by the following
parametric equation given by Saffman and Taylor in their pioneering work in 1958
\cite{st58}, 
\begin{equation}
    \label{eqn_saffman}
    \displaystyle
    z(x) = \frac{W(1-\lambda)}{2\pi}\ln\left[\frac{1}{2}\left(1+\cos\frac{2\pi x}{\lambda W}\right)\right]
\end{equation}
where $z$ is the direction of the overall front propagation, $W$ is the width of
the flow cell, and the parameter $\lambda$ is the ratio of the width of the
finger to the width of the channel. Note that this theory does not prescribe
$\lambda$ and that the surface tension between the two fluids does not enter
this equation, as it was not taken into account in its derivation. However,
Saffman and Taylor experimentally measured values of $\lambda$ close to $0.5$
over a wide range of flow rates, and it was later found that the finger width
was selected by surface tension \cite{ms81}. When the flow takes place inside a
porous medium, on the other hand, for example a Hele-Shaw cell randomly filled
with glass beads, the fingers exhibit a fractal structure \cite{f88}. Properties
of these fractal fingers are controlled by the competition between viscous and
capillary forces, the ratio of which is called the capillary number, and the
viscosity contrast of the two fluids \cite{ltz88,lz89,ga19}. For slow injection,
the process is controlled by the disorder in the capillary forces at the pores,
and the displacement front generates {\it capillary fingers} with a fractal
dimension similar to invasion percolation clusters \cite{ww83}. For fast
injection of a lower-viscous fluid into a higher-viscous fluid inside a porous
medium \cite{h87}, the growth of the front is governed by viscous instabilities,
and it produces fractal viscous fingers with a lower fractal dimension
\cite{mfj85,par21}. It was pointed out that the statistical properties of
viscous fingers in porous media are analogous to diffusion limited aggregation
(DLA) \cite{acg89}, which is a process of aggregation of matter limited by the
diffusion of random walkers arriving from a far distance with a steady flux
\cite{ws81}. Both of Saffman-Taylor viscous fingering and DLA follow Laplacian
growth,
\begin{equation} 
    \label{eqn_laplace}
    \displaystyle
    \nabla^2 p = 0
\end{equation}
where $p$ is the probability density of the random walker for DLA or the
pressure drop across the interface between the two fluids for viscous fingering
\cite{ws83,p84,lmt04}. For Saffman-Taylor viscous fingering, the flow in each of
the fluid phases is governed by Darcy's law \cite{d56,w86Darcy}, 
\begin{equation}
    \label{eqn_darcy}
    \displaystyle
    \boldsymbol v = \frac{\kappa}{\mu}\boldsymbol \nabla p\; ,
\end{equation}
where $ \boldsymbol v$ is the fluid velocity, while $\kappa$ and $\mu$ are
respectively the permeability of the porous medium and the fluid's viscosity.
Taking into account the flow incompressibility, $\boldsymbol \nabla\cdot
\boldsymbol v=0$ for incompressible fluids, the Laplace equation $\nabla P^2$
($P$ being the pressure field) holds everywhere in both fluids. It follows from
this that the same equation \ref{eqn_laplace} holds for the pressure drop across
the interface, and thus the interface displacement is driven by Laplacian growth
if the fluids' viscosities are different.

The linearity between the pressure drop and the velocity or the flow rate
indicated by Eq. \ref{eqn_darcy} does not, however, hold for {\it steady-state}
flow in a certain range of capillary numbers. Steady state flow implies the
simultaneous flow of both fluids in the porous medium for a sufficiently long
time so that the system has reached a state when the statistical averages of
macroscopic quantities, such as the saturation or the global pressure drop, do
not drift with time anymore while both fluids are still flowing. Experiments in
a porous medium consisting of a mono-layer of glass beads in which air and
water-glycerol mixture were displaced simultaneously have shown that the total
flow rate $Q$ in steady state varied with the applied pressure drop $\Delta P$
as a non-linear power law of exponent $\approx 1.85$ \cite{tkr09,tlk09}. Later
experiments with two incompressible fluids in the same porous medium found the
exponent to be $\approx 1.35$ or $1.5$ depending on the fractional flow
\cite{aet14}. Experiments with three-dimensional porous media made of glass bead
packings \cite{rcs11,rcs14,sbd17} and real core samples
\cite{glb20,zbg21,zbb22}, performed by different groups, have further
established this non-linearity with different exponent values. Various
theoretical approaches \cite{tkr09,tlk09,sh12,shb13}, and numerical modeling
with variable-radii tubes \cite{cfh23}, capillary bundles \cite{rhs19},
pore-network models \cite{sh12,sbd17} and lattice Boltzmann simulations
\cite{yts13} have been carried out to understand the origin of the non-linearity
that makes the rheology to deviate from Eq. \ref{eqn_darcy}. In general, the
non-linear power law can be expressed by the expression
\begin{equation}
    \label{eqn_steady} 
    \displaystyle
    Q \propto (\Delta P-P_{\rm t})^\beta~,
\end{equation}
where $P_{\rm t}$ is a threshold pressure that may exist so that there will be
no flow in the system below $P_{\rm t}$, and $\beta>1$ is the non-linear
exponent. Fundamentally, this non-linearity is related to the distribution of
the capillary barriers at the pore throats, which result from surface tension at
the menisci between the two fluids; disorder in the pore geometry induces
disorder in the spatial distribution of the capillary barriers. With the
increase in $\Delta P$, more and more pores progressively allow for the barrier
to be overcome and flow to be conducted, which makes $Q$ increase faster than
the increase in $\Delta P$, and hence $\beta > 1$ \cite{rh87}. When all the
available pores along the interface, at any time, start flowing at a
sufficiently high $\Delta P$, the relation becomes linear. For a simplified
porous system such as a capillary bundle, the value of $\beta$ can be determined
analytically by integrating the individual flow rates of the tubes over the
distribution of the capillary barriers \cite{rhs19}. For a porous network, the
distribution of pore throat sizes \cite{rsh21} and the wettabilities
\cite{fsr21,fsh23} strongly control the value of $\beta$, as the distribution of
capillary barriers depends on them.

In this article, we investigate to which extent this non-linearity in the flow
rate, which has been demonstrated to exist for the steady state due to disorder
in the capillary barriers, also exists for the growth of viscously-unstable
fingers during drainage, and what are its characteristics. The Eq.
\ref{eqn_laplace} for DLA considers $p=0$ at the perimeter of the aggregate,
which, fingering in two-phase flow, would correspond to the absence of surface
tension. The analogy to DLA therefore only applies for capillary numbers that
are sufficiently high for capillary forces to be insignificant as compared to
viscous forces. In the intermediate regime between capillary and viscous
fingerings, we may expect that capillary forces will compete with viscous
forces, so that the disorder in capillary barriers will play a role in the
finger growth. L{\o}voll et al \cite{lmt04} and Toussaint et al \cite{tlm05}
have investigated this displacement regime experimentally using two-dimensional
(2D) porous media consisting of Hele-Shaw cells filled with a monolayer of glass
beads, and measured the fractal dimension of the fingers and the statistical
width of the front \cite{acg89,aac91,aet96}, comparable to $\lambda$ in Eq.
\ref{eqn_saffman}. Based on these measurements, they showed that the drainage
fingers are characteristically different from DLA. They further proposed a {\it
quadratic} relationship between the finger growth rate and the local pressure
gradient in the intermediate flow regime, resulting from pore-scale disorder.
For DLA however, an ad-hoc type surface tension was considered  by introducing a
sticking probability \cite{ws83}, which showed no change in the fractal
dimension \cite{tnk88,ntl90}. It was then argued that the two-phase flow fingers
are more similar to dielectric breakdown models (DBM) (or the $\eta$ models)
\cite{npw84,a89,mj02}, which are generalizations of DLAs with power-law
relationships of exponent $\eta$ between the growth probability and the local
growth/displacement-driving field.

The experimental study on 2D porous media \cite{lmt04,tlm05} left a number of
open questions, in particular concerning the dependence of the nonlinear growth
exponent and statistical width of the viscous fingers on the capillary number.
In this paper we study viscously-unstable drainage as a function of the
capillary number using large-scale simulations in dynamic pore networks. We
inject a lower-viscosity fluid at one edge of the network filled with a higher
viscosity fluid, and characterize the statistical properties of the resulting
viscous finger in a reference frame attached to the most advanced finger tip, as
the finger grows inside the porous medium. We explore in particular the
relationship between the finger's growth rate and the local pressure drop across
the interface, and how it depends on the capillary number. We also focus on the
statistical width parameter $\lambda$, analog to the $\lambda$ in Eq.
\ref{eqn_saffman}, compare it with the values obtained in the previous
experimental study as well as in studies of DLA and Saffman-Taylor viscous
fingering, and characterize its dependence on the capillary number. 

In the following, we first describe the computational model and the simulation
procedure in Sec. \ref{sec_simulation}, which is based on dynamic pore-network
modeling \cite{sgv21} with a specific boundary condition. In Sec.
\ref{sec_results} we present our results, where we first characterize the
statistical profiles related to the volume and growth of the fingers in
subsection \ref{sec_vol}. In subsection \ref{sec_shape} we characterize the
shape profile and measure the width ratio $\lambda$. In subsection
\ref{sec_pressure} we then explore the relationship between the growth rate and
the local pressure drop across the interface. Finally, we discuss the results
and provide an overall conclusion in Sec. \ref{sec_summary}. 

\section{System description and modeling}
\label{sec_simulation}

\begin{figure}
    \centerline{\hfill\includegraphics[width=0.4\textwidth]{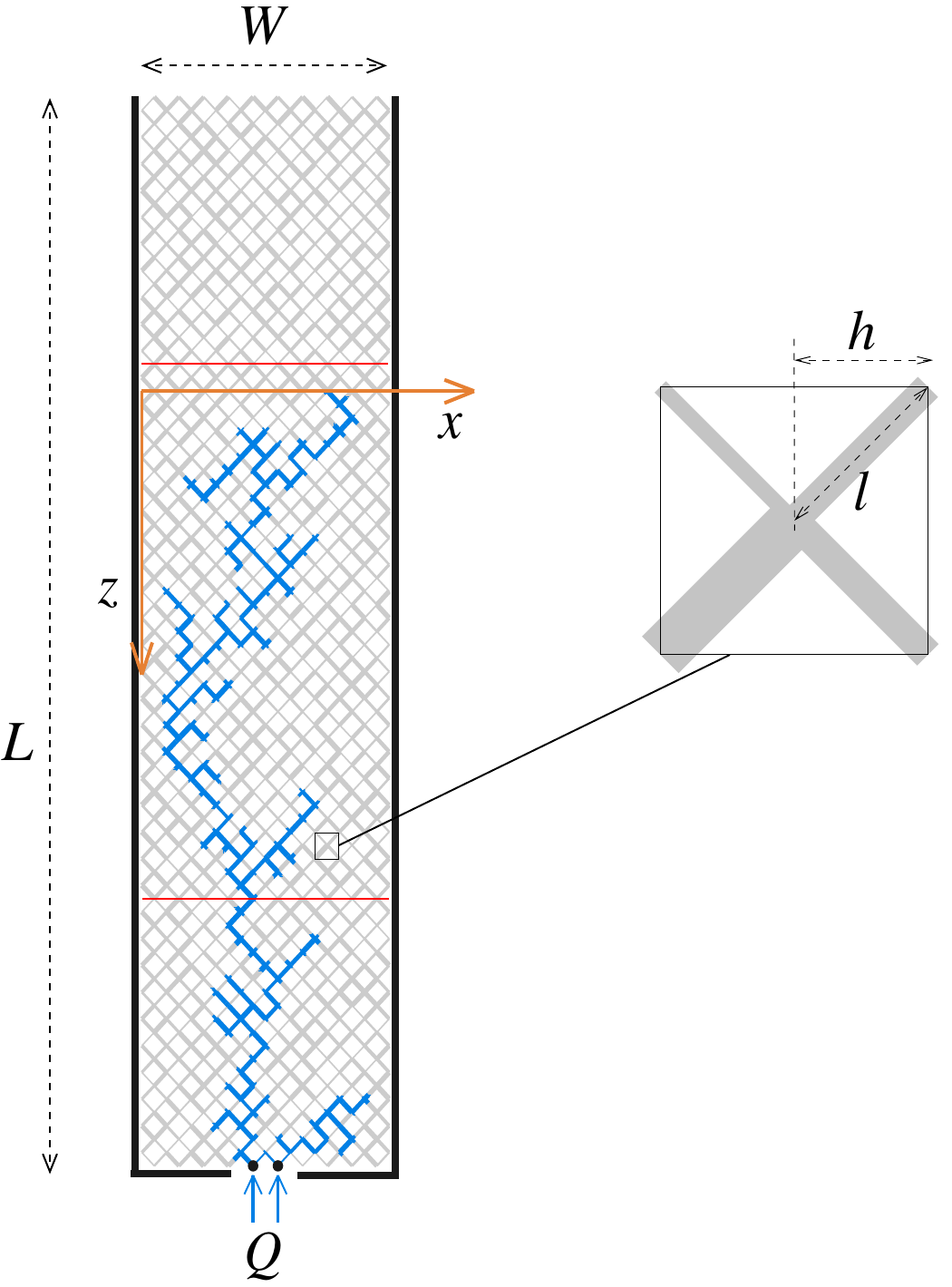}\hfill}
    \caption{\label{fig_network} A pore-network of dimension $L\times W$
    consisting of $N_W=20$ and $N_L=80$ links. The length of each link is $l$
    and therefore $h=l/\sqrt{2}$. The network was initially filled with
    high-viscosity fluid (represented in gray). The low-viscosity fluid
    (represented in blue) is injected with a total flow rate $Q$ through $N_I=2$
    inlet nodes at the bottom, indicated by the black dots. All the other nodes
    at the bottom boundary, and all on the two vertical boundaries are closed,
    which is indicated by the thick black lines. The nodes at top boundary are
    open and work as the outlets. The $(x,z)$ coordinate axes indicated by the
    orange lines are attached to the most advanced fingertip and move with it.
    The part of the network between the two red horizontal lines is used for the
    measurements of different quantities in order to avoid finite size effects
    at the inlet and the outlet.}
\end{figure}

\begin{figure}
\centerline{
     \subfloat[\label{fig_model-a}]{\includegraphics[width=0.20\textwidth]{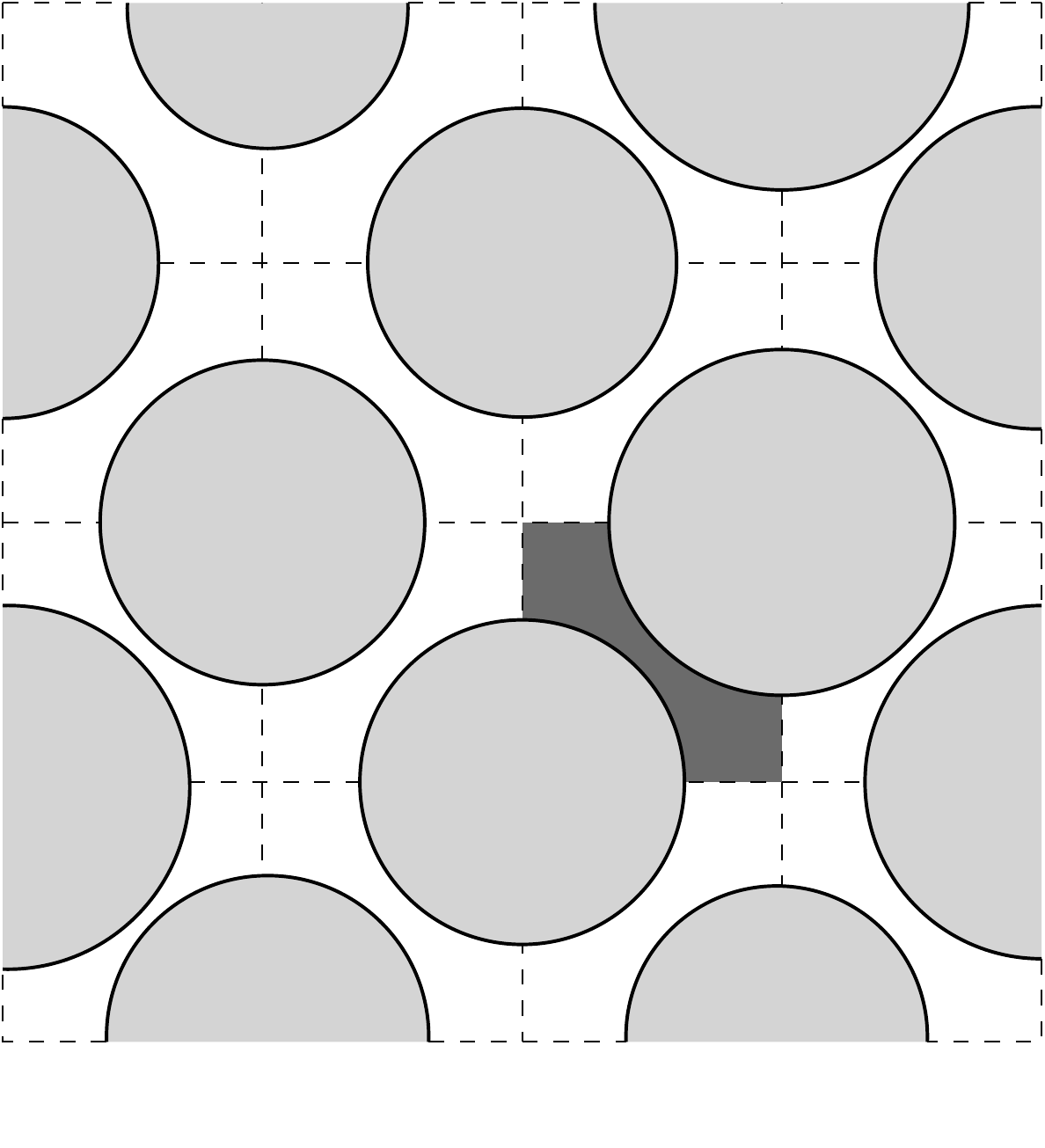}}\hfill
     \subfloat[\label{fig_model-b}]{\includegraphics[width=0.26\textwidth]{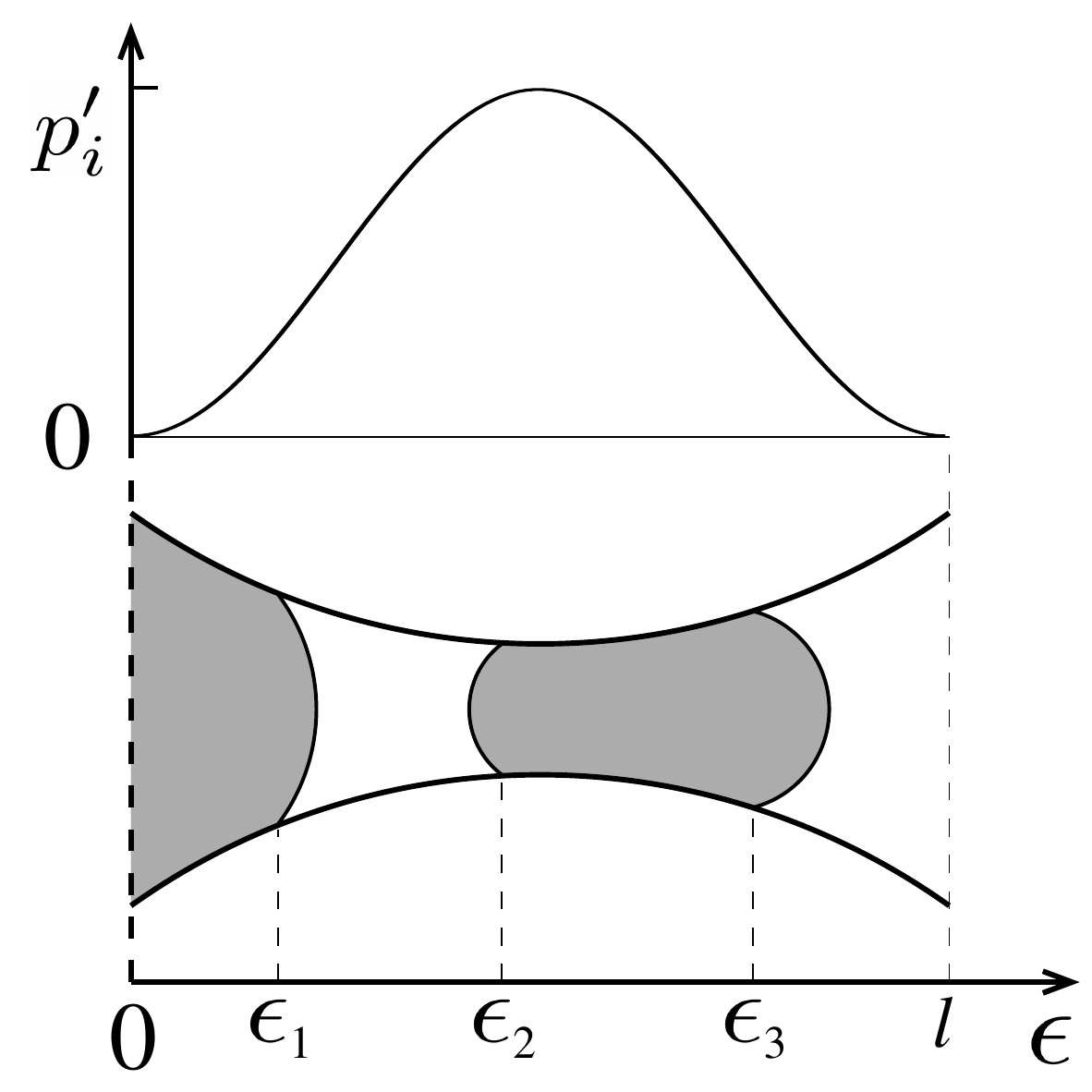}}}
     \caption{\label{fig_model} (a) Illustration of the network of pores and
     nodes. The pore space is indicated by white color whereas the gray circles
     represent the grains in a porous rock or the glass beads in a Hele-Shaw
     cell. One of the links is colored by dark gray. The links intersect at
     nodes at the intersections  of the dashed lines. (b) Shape of individual
     pores in the mid-horizontal plane of the network, and variation of the
     capillary pressure $p'$, given by Eq. \ref{eqn_young}, along the length of
     the pore. The white and gray segments inside the pore represent
     respectively the wetting and non-wetting fluid blobs; there are three
     interfaces between them in this case, which are in contact with the beads
     at positions $\epsilon_1$, $\epsilon_2$ and $\epsilon_3$ along the pore
     length.}
\end{figure}

The core of our simulation consists of a dynamic pore-network model
\cite{sgv21}, which we tweaked to adapt to the present problem. The model
has been developed for over a decade \cite{amh98} and tested against numerous
experimental \cite{est13,sbd17}, theoretical \cite{sh12,rsh20,rps22,fsh23b} and
lattice-Boltzmann simulations \cite{sgv19}. In this computational method the
pore space of a porous medium is modeled by a network of links and nodes. Our
network is spread in 2-dimensions (2D) within a flat cuboid domain similar to a
Hele-Shaw cell, and consists of $N_W\times N_L$ number of links embedded in a
diamond lattice. Such a network with $N_W=20$ and $N_L=80$ is shown in Fig.
\ref{fig_network}. Here the subscripts $W$ and $L$ refer to the directions of
the network orthogonal and parallel to the overall flow respectively. Each link
of the network is of length $l=1\,\text{mm}$. The total network is therefore
$W=hN_W\,\text{mm}$ wide and $L=hN_L\,\text{mm}$ long, where $h=l/\sqrt{2}$. We
assign the entire pore-space of the network to the links, so the nodes only
represent the position of the link intersections. The correspondence between
this geometry and that of a granular quasi-two-dimensional porous medium
consisting of cylinders is illustrated in Fig. \ref{fig_model-a}, where a small
part of the porous medium is shown. The links are therefore composite in nature,
which means that each link must represent a narrow pore throat in between two
wider pore bodies. Such a geometry is modeled by an hourglass-shaped
converging-diverging tube (as illustrated in Fig. \ref{fig_model-b}) whose
cross-sectional area varies along its length. This results in a variation in the
capillary pressure $p'$ as the interface (i.e., meniscus) between the two fluids
moves along a link. We model this variation with a modified Young-Laplace
equation \cite{d92,amh98,shb13},
\begin{equation}
  \displaystyle
  |p'_i(\epsilon)| = \frac{2\gamma}{r_i}\left[1-\cos\left(\frac{2\pi \epsilon}{l}\right)\right] \;,
  \label{eqn_young}
\end{equation}
where $y$ is the position of an interface inside a link $i$. Such variation in
the capillary pressure is shown in Fig. \ref{fig_model-b}. Here
$\gamma=\gamma'\cos\theta$, where $\gamma'$ and $\theta$ represent the surface
tension between the two fluids and the contact angle respectively. The $r_i$ is
the average radius of the $i$th link, the value of which we choose from a
uniform distribution of random numbers in the range between $0.1l$ to $0.4l$,
which introduces the disorder in the network.

The capillary numbers for both the viscous and capillary fingering regimes are
sufficiently small for any inertial effect to be negligible. For fully developed
laminar flow of incompressible fluids in each link, we therefore consider the
following equation for the flow rate of the fluids \cite{w21,ltz88},
\begin{equation}
  \displaystyle
  q_i = -\frac{k_i}{l\mu_i}\left[\Delta p_i -\sum_b p'_i(\epsilon_b)\right]~, 
  \label{eqn_washburn}
\end{equation}
where $\Delta p_i$ is the pressure drop between the two nodes across the $i$th
link. The network is assumed to be placed horizontally and no gravity is
considered. Here $\mu_i$ is the effective viscosity of the tho fluids inside the
link, which is given by $\mu_i=s_i\mu_{\rm n}+(1-s_i)\mu_{\rm w}$, where
$\mu_{\rm n}$ and $\mu_{\rm w}$ are the viscosities of the non-wetting and
wetting fluids respectively and $s_i$ is the non-wetting saturation in the $i$th
link, i.e., the proportion of the link volume that is occupied by the
non-wetting fluid. The term $k_i$ represents the mobility of the link, given by
$k_i=a_ir_i^2/8$ for Hagen-Poiseuille flow in a circular cross-section where
$a_i=\pi r_i^2$ is the cross-sectional area of the link. The summation over the
capillary pressure $p'_i$ in Eq. \ref{eqn_washburn} is over all the interfaces
inside the link $i$, obtained using Eq. \ref{eqn_young}. For the link shown in
Fig. \ref{fig_model-b}, for example, the summation will be over the three
interfaces at $\epsilon_1$, $\epsilon_2$ and $\epsilon_3$.

To find out the pressures $p_j$ at every node $j$ of the network, we use the
Kirchhoff equation for incompressible flow $\sum_{i\in n_j} q_i=0$ for each
node. Here the summation is over the links $n_j$ connected to a node $j$. This
provides a set of linear equations which we solve by conjugate gradient method
\cite{bh88} with proper boundary conditions. This is illustrated in Fig.
\ref{fig_network} where the two nodes ($N_I=2$) at the bottom edge act as the
inlets. All the nodes at the top edge work as the outlets through which fluids
leave the system. The two vertical boundaries of the network parallel to the
overall flow are closed. The network is initially filled with wetting fluid
(gray colored in the figures), and we inject the non-wetting fluid (blue colored
in the figures) through inlet nodes with a total constant flow rate $Q$. We
therefore set $p_j=0$ for all the outlet nodes and $q_i=Q/N_I$ for all the
virtual links connected to the inlet nodes, indicated by the two arrows in Fig.
\ref{fig_network}. These serve as the boundary conditions for solving the linear
system of equations for the node pressures.

The positions of the two fluids inside every link are assigned by the positions
of the interfaces between the blobs of the two immiscible fluids in every link
as shown in Fig. \ref{fig_model-b}. At every time step, the displacements of the
two fluids are performed by updating these positions by a distance,
\begin{equation}
  \displaystyle
   \Delta \epsilon_i=\Delta t\, q_i/a_i
  \label{eqn_timestep}
\end{equation}
in the direction of the flow in the corresponding link. Here $\Delta t$ is a
time step chosen in such a way that the displacement of an interface inside any
link do not exceed more than $0.1l$.

One final detail about the modeling is how to distribute the fluids from links
to their neighboring links through the nodes. For this, first the links carrying
fluids towards and away from every node are identified, we call them the
incoming and outgoing links, respectively. Then for every node $j$, the total
volumes of the wetting and non-wetting fluids ($V^{\rm w}_j$ and $V^{\rm n}_j$
respectively, and $V_j=V^{\rm w}_j+V^{\rm n}_j$) arriving from the incoming
links to the node are measured using Eqn. \ref{eqn_timestep}. These volumes are
then distributed to each outgoing links $i$ by placing new wetting and
non-wetting blobs of volumes $V^{\rm w}_i=q_i\Delta t V^{\rm w}_j/V_j$ and
$V^{\rm n}_i=q_i\Delta t V^{\rm n}_j/V_j$ respectively. This means that the
ratio of $V^{\rm w}_i$ to $V^{\rm n}_i$ in any outgoing link is identical to the
ratio between the incoming wetting and non-wetting volumes at the corresponding
node, and the ratio between the total volumes ($V^{\rm w}_i+V^{\rm n}_i$)
injected in different outgoing links $i$ is the same as the ratio between the
flow-rates $q_i$ in those links. Further technical details about this algorithm
can be found in \cite{sgv21}.

\section{Results and discussions}
\label{sec_results}

\begin{figure*}[htbp]
    \captionsetup[subfigure]{labelformat=empty}
    \centerline{
        \subfloat[$\mathrm{Ca}=0.004$]{\includegraphics[width=0.1\textwidth]{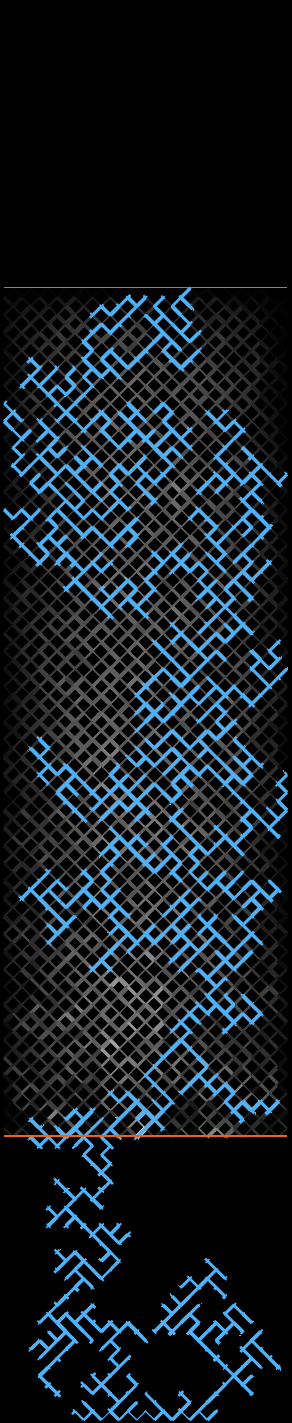}}\hfill
        \subfloat[$\mathrm{Ca}=0.008$]{\includegraphics[width=0.1\textwidth]{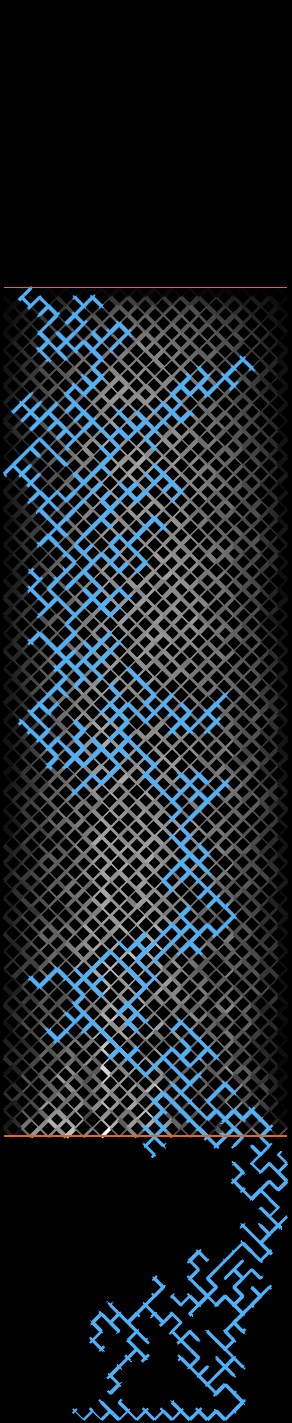}}\hfill
        \subfloat[$\mathrm{Ca}=0.01$ ]{\includegraphics[width=0.1\textwidth]{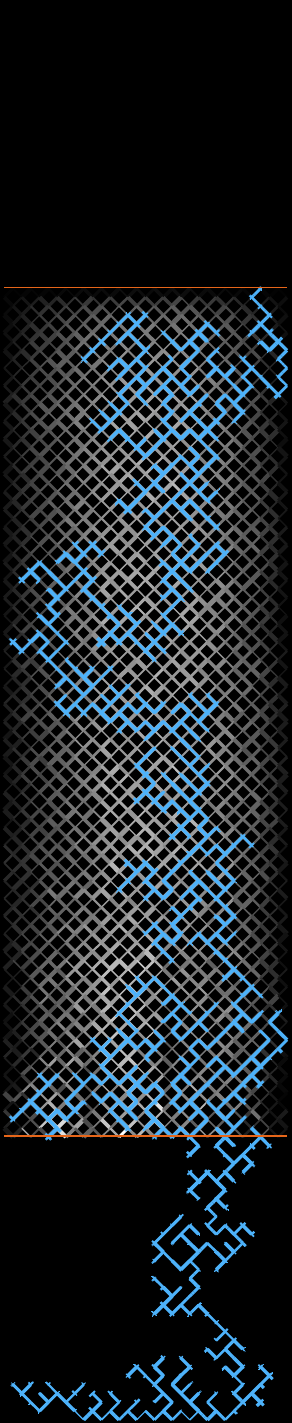}}\hfill
        \subfloat[$\mathrm{Ca}=0.04$ ]{\includegraphics[width=0.1\textwidth]{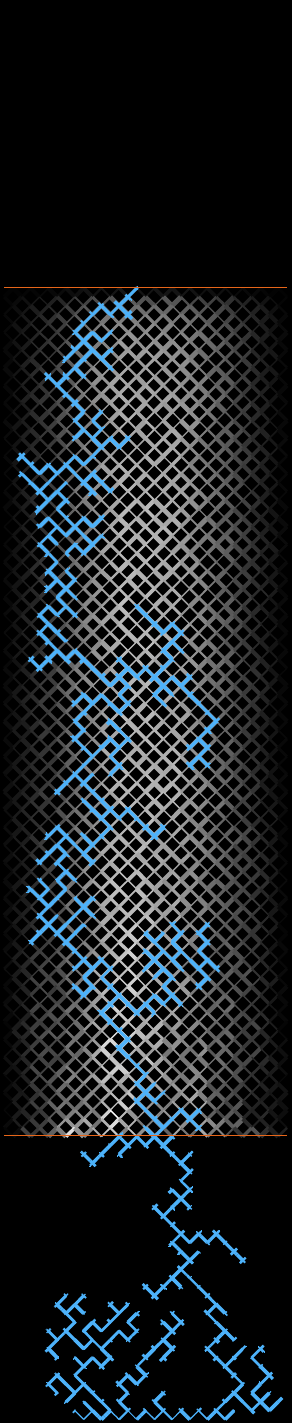}}\hfill
        \subfloat[$\mathrm{Ca}=0.08$ ]{\includegraphics[width=0.1\textwidth]{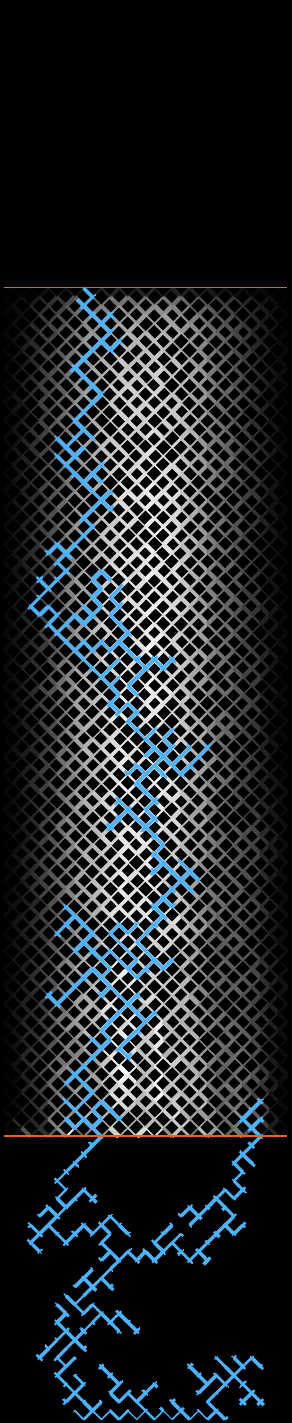}}\hfill
        \subfloat[$\mathrm{Ca}=0.1$  ]{\includegraphics[width=0.1\textwidth]{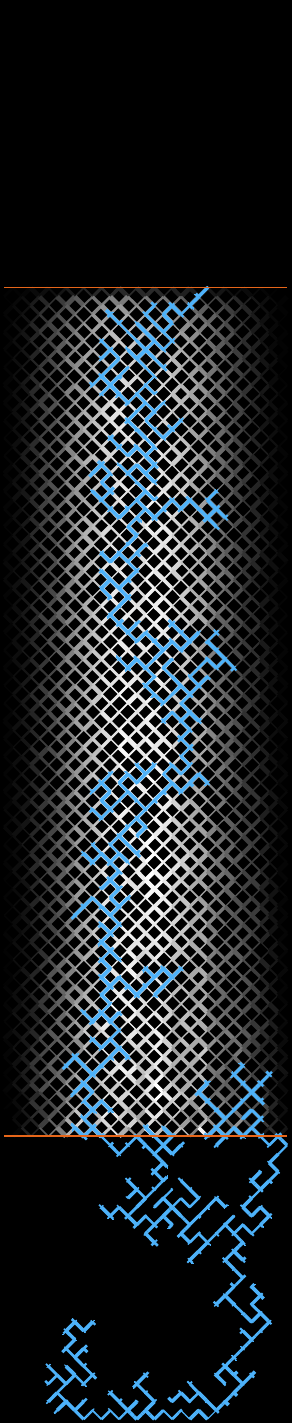}}\hfill
        \subfloat[$\mathrm{Ca}=0.4$  ]{\includegraphics[width=0.1\textwidth]{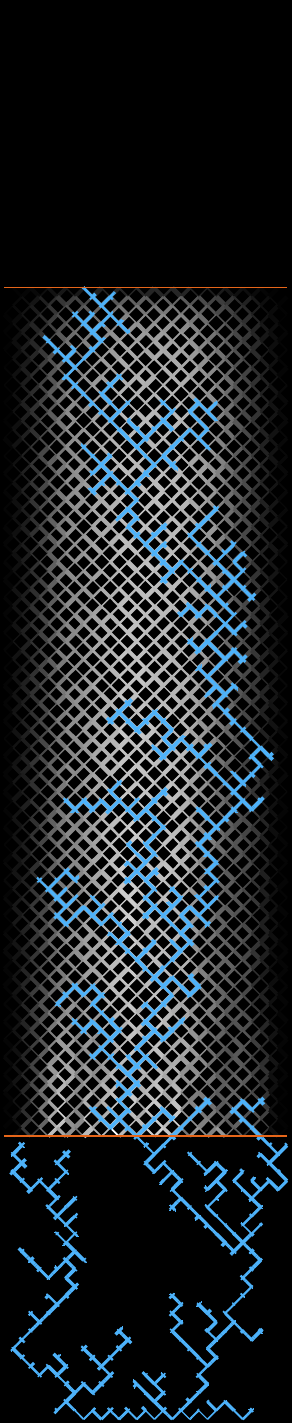}}\hfill
        \subfloat[$\mathrm{Ca}=0.8$  ]{\includegraphics[width=0.1\textwidth]{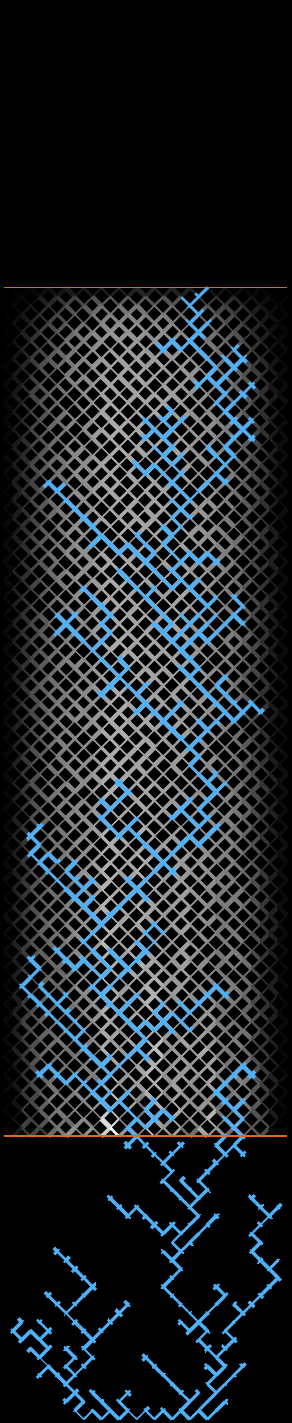}}}

    \caption{\label{fig_snapshots} Drainage fingers generated for viscosity
    ratio $M=10^{-4}$, and for different values of Ca. The blue and black colors
    respectively denote the invading non-wetting and the defending wetting
    fluids. The inlets are at the bottom edge of the networks and the overall
    flow direction is towards the top. The gray shade indicates the non-wetting
    volume density $\tilde\sigma(x,z)$, averaged over time and samples in the
    moving coordinate system.}  
\end{figure*}

Drainage simulations are performed at constant flow rates $Q$ corresponding to
chosen the capillary numbers Ca. This is a dimensionless global parameter of the
flow, which is defined as the average ratio of the viscous forces to the
capillary forces at the pore level. It is defined as \cite{ltz88},
 \begin{equation}
  \displaystyle
   \text{Ca} = \frac{\mu Q/A}{\gamma} 
  \label{eqn_casteady}
\end{equation}
where $A$ is the cross-sectional area. Here we point out that for our
pore-network only the areas of the channels are properly defined, so $A$ is the
orthogonal cross-section of the channels alone, and therefore $Q/A$ is the true
mean longitudinal fluid velocity, rather than the filtration (or Darcy) velocity
as usually considered in experiments. Note also that experimental studies,
including those addressing the same topic \cite{lmt04,tlm05} may have used a
different definition of Ca, considering in particular that the viscous pressure
drop should involve the medium's permeability; Eq. \ref{eqn_casteady} assumes
that the permeability can be equated to the square of the typical channel
cross-section. Furthermore, the wetting angle could be taken into account in the
capillary pressure drop, thus appearing in the definition of Ca as well. When
comparing our simulations results to experimental results under identical flow
conditions, the similarity of the flow conditions must be assessed based on Ca
estimates obtained from the same definition, so a conversion of the Ca values
(either experimental, or numerical) may be necessary.

To find out the value of $Q$ that must be imposed in the simulation for a given
Ca, we first define it as \cite{lmm05},
\begin{equation}
  \displaystyle
  {\rm Ca} = \frac{\Delta p_{\rm visc}}{\Delta p_{\rm cap}} = \frac{\Delta P_Wh/L}{2\gamma/{\bar r}}
  \label{eqcasim}
\end{equation}
where $\Delta p_{\rm visc} = \Delta P_Wh/L$ and $\Delta p_{\rm cap} =
2\gamma/{\bar r}$ are respectively the average viscous and capillary pressure
drops across a pore. Here $\Delta P_W$ is the total viscous pressure drop across
the network when it is completely filled with the wetting fluid and ${\bar r}$
is the average link radius. We then first solve the viscous pressure drop
$\Delta P_{{\rm visc},1}$ for a test flow rate $Q_1$ for the network saturated
with wetting fluid, and then determine the flow rate $Q$ by using,
\begin{equation}
  \displaystyle
  Q = \frac{2\gamma L}{h{\bar r}}\frac{Q_1}{\Delta P_{{\rm visc},1}}{\rm Ca} \;.
  \label{eqca2q}
\end{equation}
This is because $Q\propto \Delta P_W$ for single phase flow. The simulation is
then performed with $q_i = Q/N_I$ for all the virtual inlet links.

\begin{table*}
    \caption{\label{tab_samples}Number of samples simulated for different values
    of Ca and $M$. The values of $M$ are $10^{-4}$ and $10^{-5}$ whereas the
    values of Ca are in the range $0.01$ to $0.9$. All the measured quantities
    are statistically averaged over these many samples for the respective values
    of Ca and $M$.}
    \begin{ruledtabular}
        \begin{tabular}{rrrrrrrrrrrrrrrrrrr}
            Ca:           & 0.01 & 0.02 & 0.03 & 0.04 & 0.05 & 0.06 & 0.07 & 0.08 & 0.09 & 0.10 & 0.20 & 0.30 & 0.40 & 0.50 & 0.60 & 0.70 & 0.80 & 0.90 \\ \hline
            $M = 10^{-4}$ &   57 &   79 &   93 &  105 &  112 &  117 &  123 &  125 &  125 &  125 &  121 &  105 &   98 &   90 &   84 &   81 &   76 &   73 \\
            $M = 10^{-5}$ &   30 &   49 &   56 &   66 &   71 &   74 &   81 &   81 &   82 &   81 &   80 &   65 &   62 &   57 &   54 &   51 &   46 &   45 \\
        \end{tabular}
    \end{ruledtabular}
\end{table*}

We have considered a network of dimension $32\times 160$ links in this study,
this is based on the maximum computational time we could spend reasonably. We
had to make additional sweeps through the entire network at every time step to
compute the volume density, the growth density and the local pressure drops,
which we will discuss later. This made the simulations considerably more
time-consuming compared to usual pore-network simulations with the same model.
Furthermore, in the analyses we discarded the regions of the network that are
distant from the inlet and the outlet by less  than $N_W$ rows, as shown by the
red lines in Fig. \ref{fig_network}, in order to avoid any boundary effect. In
Tab. \ref{tab_samples} we show the numbers of different realizations of the
network considered for different values of $M$ and Ca. All the measured
quantities in different plots in this article are averaged over these many
samples. The surface tension between the fluids was chosen to
$\gamma=0.03\,\text{N/m}$. Two viscosity ratios, $M=\mu_{\rm n}/\mu_{\rm w} =
10^{-4}$ with $\mu_{\rm n}=10^{-5}\,\text{Pa.s}$ and $\mu_{\rm
w}=10^{-1}\,\text{Pa.s}$, and $M=10^{-5}$ with $\mu_{\rm
n}=10^{-5}\,\text{Pa.s}$ and $\mu_{\rm w}=1\,\text{Pa.s}$, were considered. A
typical set of simulations for different capillary numbers and viscosity ratios
are shown in Fig. \ref{fig_snapshots}, where the black and blue colors represent
the wetting and non-wetting fluids respectively. The direction of the overall
flow in these images is from the bottom to the top as indicated by the arrow.
The two red lines indicate the aforementioned regions near the inlets and the
outlets that are not taken into account in the analyses. The gray shades
indicate the average volume density, which we describe in the following.

\begin{figure*}[htbp]
    \centerline{\hfill
    \includegraphics[width=0.45\textwidth]{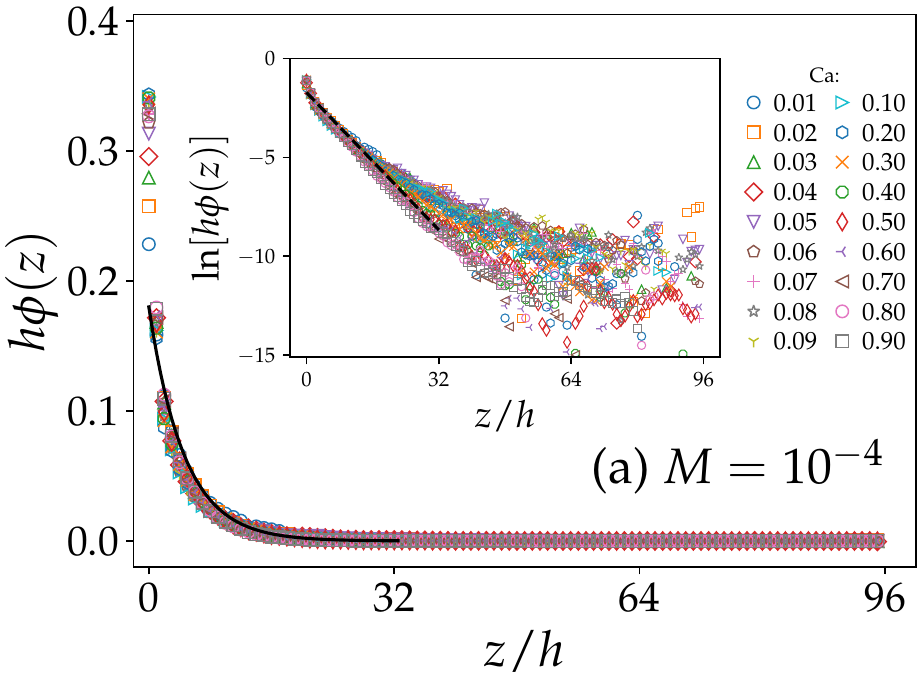}\hfill
    \includegraphics[width=0.45\textwidth]{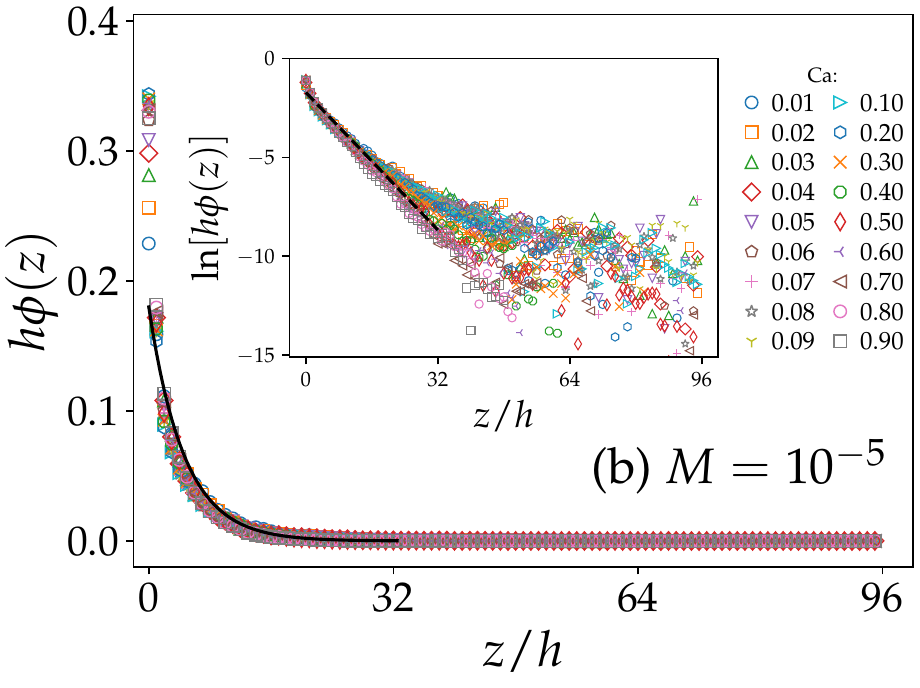}\hfill}
    \caption{\label{fig_growthden} Plot of the dimensionless growth density
    $h\phi(z)$ as a function of the row numbers from the tip, $z/h$, for the
    viscosity ratios (a) $M=10^{-4}$ and (b) $M=10^{-5}$. The different sets in
    each plot correspond to different values of Ca indicated in the legend of
    both plots. The solid line corresponds the model function
    $B\exp(-\frac{z}{h\xi})$ with $\xi = 4.6$ and $B=0.18$, for both plots. The
    inset shows a linear-logarithmic plot of the same data where the dashed line
    has a slope of $-1/\xi$ and intercept of $\ln B$.}
\end{figure*}

\subsection{Longitudinal profiles of volume density and volumetric growth rate}
\label{sec_vol}
The volume density of the fingers represents a statistical map of the occupation
of the two-dimensional network space by the invading non-wetting fluid. This
provides a statistical {\it shape} of the displacement structure which is
comparable with the smooth continuum-medium fingers \cite{st58}. The density
maps are functions of the distance from the most advanced fingertip, $z$, as the
displacement process can be considered statistically stationary in an $(x,z)$
reference frame attached to that tip (see figure \ref{fig_network}), as
previously shown for experimental drainage fingers \cite{lmt04,tlm05}, as well
as for DLA fingers \cite{acg89,aac91}. With respect to this dynamic $(x,z)$
reference frame, we define the {\it volume density} $\tilde\sigma(x,z)$ as,
\begin{equation}
    \displaystyle
    \tilde\sigma(x,z) = \frac{1}{h^2n(x,z)}\sum_{k=1}^{n(x,z)} V_k(x,z) \;,
    \label{eqn_volden}
\end{equation}
which is essentially the volume of  injected fluid per unit area at a given
position in the $(x,z)$ reference plane, averaged over time, $V_k(x,z)$ being
the non-wetting volume in the link positioned at position ($x,z$) at time step
$k$. Here we point out that the lowest unit of resolution for our measurements
is one single link, which means that we have only one data point within an area
of $h^2$. Furthermore, as the available window of network for the measurements
varies with time, the number of elements withing the summation is different for
different links. This is accounted by $n(x,z)$ in the above expression, which is
the number of times the volume in the link at $(x,z)$ is added in the summation.
This also means that the data near the fingertip contain better statistical
averages compared to those far behind the tip. The map is then further averaged
over a number of different samples of the network as indicated in Tab.
\ref{tab_samples} to obtain statistically averaged $\tilde\sigma(x,z)$ map. In
Fig. \ref{fig_snapshots}, we show the spatial distribution of
$\tilde\sigma(x,z)$ as gray-scale maps where the dark and light shades
correspond to the smallest and largest values, respectively. For an observer
sitting at the origin of the moving coordinate system, this volume profile of
the finger will remain the same statistically during the invasion of the fingers
in an infinitely long network, and only the length of the profile will increase
in the $z$ direction.

We next measure the volumetric {\it growth rate} $\tilde \nu(x,z)$ of the
invading fingers, which is defined as the average increase in the volume of the
finger per unit time and unit area at $(x,z)$
\begin{equation}
    \displaystyle
    \tilde \nu(x,z) = \frac{1}{h^2n(x,z)}\sum_{k=1}^{n(x,z)} \frac{1}{\Delta t_k}\left[V_k(x,z)-V_{k-1}(x,z)\right] \;,
    \label{eqn_dvol}
\end{equation}
where $\Delta t_k$ is the time interval between the time steps $k$  and $k-1$.
Note that, $\tilde \nu(x,z)$ has the dimension of a velocity. By integrating
$\tilde\nu(x,z)$ in both $x$ and $z$ directions we have
\begin{equation}
    \displaystyle
    \int_0^L\int_0^W \nu(x,z)\dif x \dif z = Q \;,
    \label{eqn_volsum}
\end{equation}
as the total increase in the volume of the whole finger per unit time is equal
to the volumetric rate of the steady fluid injection into the system. By
normalizing $\tilde\nu(x,z)$ by $Q$, we then define the {\it growth density}
$\tilde\phi(x,z)$, 
\begin{equation}
    \displaystyle
    \tilde\phi(x,z) = \frac{1}{Q}\tilde\nu(x,z) \;.  
    \label{eqn_growthden}
\end{equation}
We assume that the system is statistically symmetric with respect to the
longitudinal mid-section ($x=w/2$) of the flow cell, and we define the
quantities $\sigma(z)$, $\nu(z)$ and $\phi(z)$ in the longitudinal direction as
a function of the sole distance $z$ from the tip, by integrating
$\tilde\sigma(x,z)$, $\tilde\nu(x,z)$ and $\tilde\phi(x,z)$ in the $x$
direction,
\begin{equation}
    \label{eqn_denz}
    \displaystyle
    \begin{aligned}
            \sigma(z) & = \int_0^W \tilde\sigma(x,z)\dif x \;,\\  
            \nu(z)    & = \int_0^W \tilde\nu(x,z)\dif x    \;,\\
            \phi(z)   & = \int_0^W \tilde\phi(x,z)\dif x   \;,
    \end{aligned}
\end{equation}
from which we can also define $\phi(z)=\nu(z)/Q$. In Fig. \ref{fig_growthden},
we plot $h\phi(z)$ as a function of the number of rows from the most advanced
tip, $z/h$, for different values of Ca and $M$. Near the fingertips, the data
shows an exponential-type decay of $\phi(z)$. This is indicated by the solid
line where we plot a model function $B\exp(-\frac{z}{h\xi})$ with $\xi = 4.6$
and $B = 0.18$. Here $\xi$ is a characteristic decay length, or screening
length, which characterizes the active invasion zone. The value of $\xi$ appears
to be the same for the whole range of capillary numbers and viscosity ratios,
showing that it is a characteristic of the porous medium, which was also
observed in the drainage experiments Hele-Shaw cell filled with a mono-layer of
glass beads\cite{lmt04}. The same data are plotted in a linear-logarithmic scale
in the insets, which show that the exponential decay is valid for $z<W/2$ for
lower capillary numbers whereas the validity increases up to $z<W$ at high
capillary numbers. This is indicated by the linear dashed line which has a slope
$-1/\xi$ and intercept $\ln(B)$. This characteristic is also in agreement with
experiments \cite{lmt04} which showed similar exponential decay of finger growth
within the same range from the tip. Away from the tip, $\phi(z)$ is negligibly
small as seen from the scattered data points in the insets, which mean that the
active growth of the fingers happen near the fingertip whereas they are almost
frozen far behind the tip.


\begin{figure*}
    \centerline{\hfill
    \includegraphics[width=0.45\textwidth]{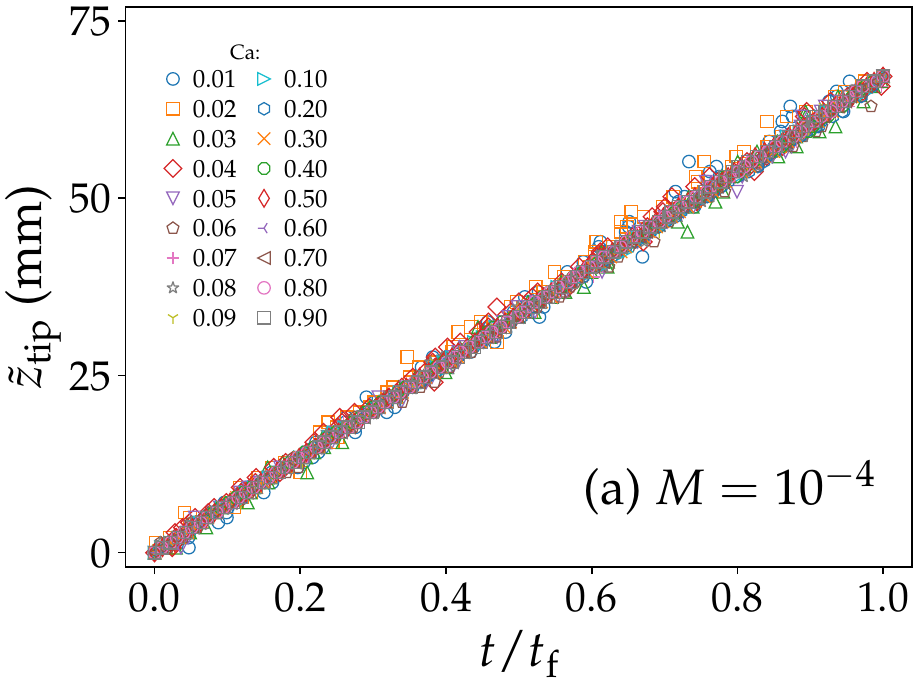}\hfill
    \includegraphics[width=0.45\textwidth]{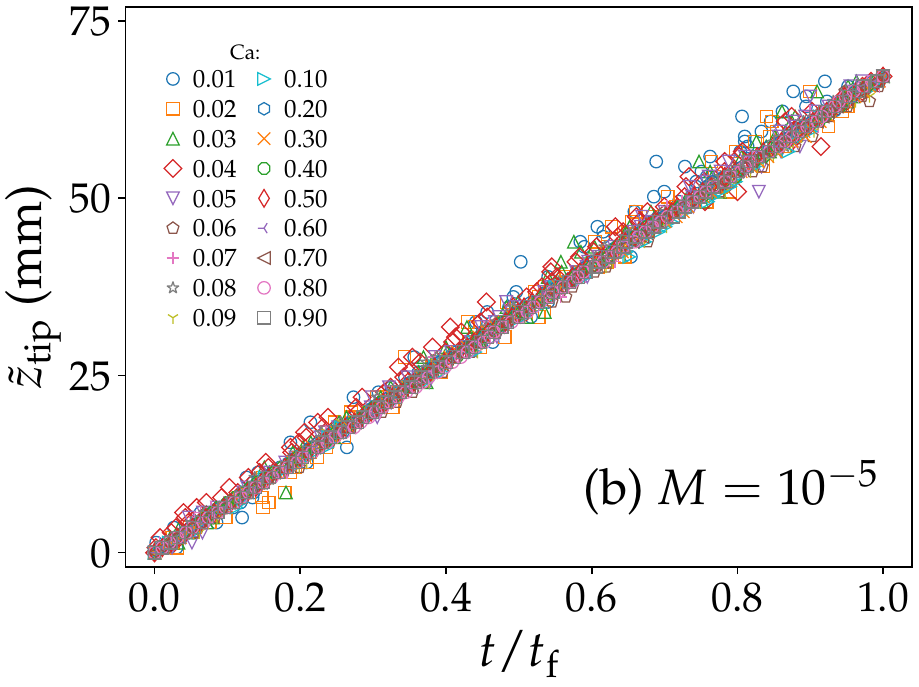}\hfill}
    \caption{\label{fig_tipvel} Plot of the position of the most advanced
    fingertip, $\tilde{z}_\text{tip}$, as a function of the scaled time
    $t/t_\mathrm{f}$ where $t_\mathrm{f}$ is the time when the most advanced
    fingertip reaches the final position in the simulation. The two plots
    correspond to the viscosity ratios (a) $M=10^{-4}$ and (b) $M=10^{-5}$; the
    different symbols correspond to the different values of Ca indicated in the
    plot legends.}
\end{figure*}

\begin{figure*}[htbp]
    \centerline{\hfill
    \includegraphics[width=0.45\textwidth]{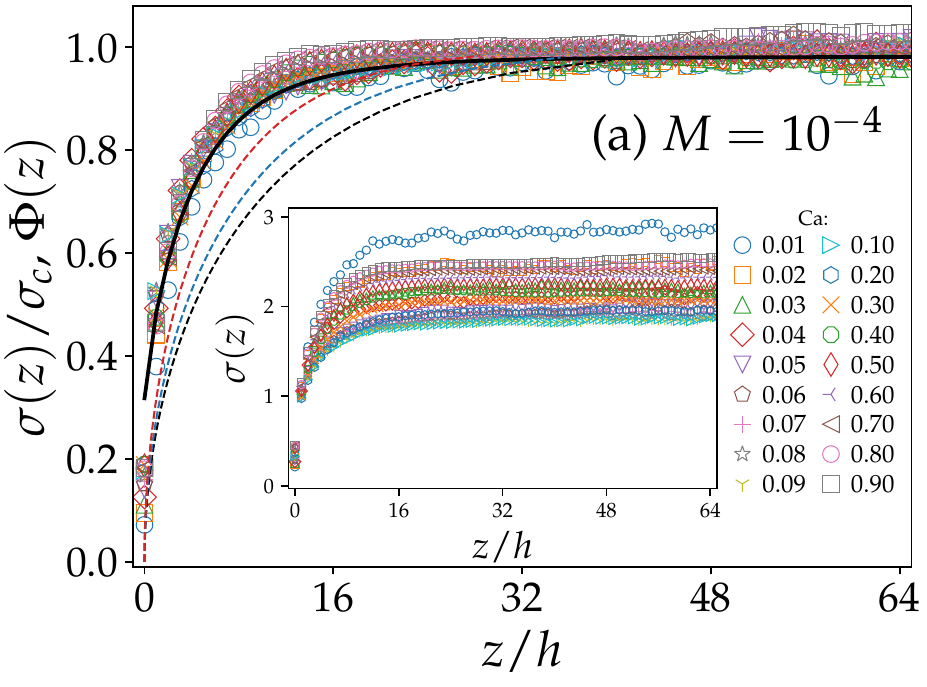}\hfill
    \includegraphics[width=0.45\textwidth]{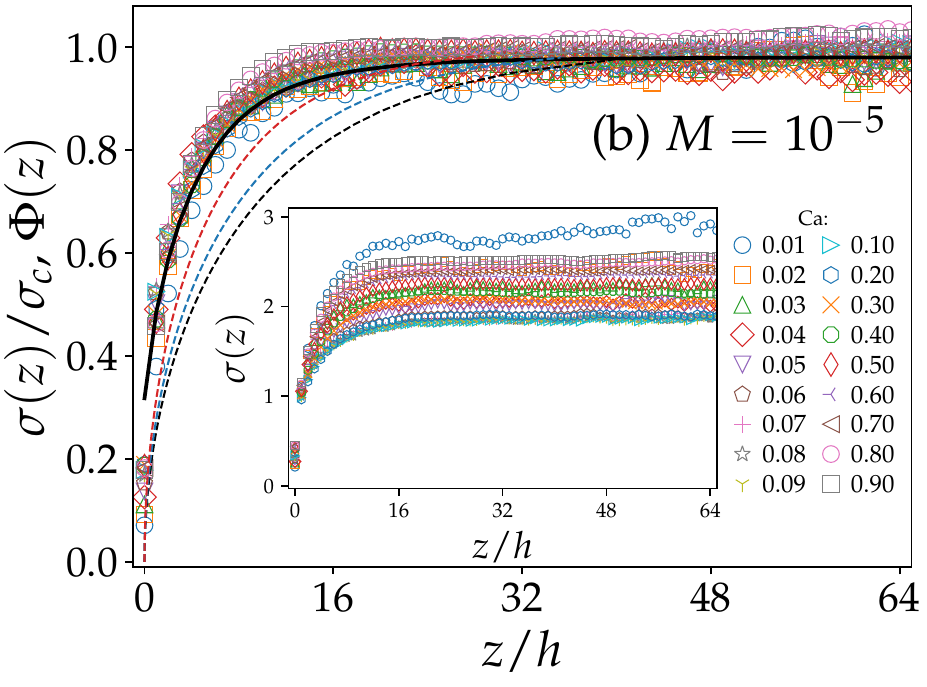}\hfill}
    \caption{\label{fig_volden} Plot of the scaled longitudinal volume density
    $\sigma(z)/\sigma_c$ as a function of the row numbers $z/h$ behind the most
    advanced fingertip. The two figures correspond the viscosity ratios (a)
    $M=10^{-4}$ and (b) $M=10^{-5}$ and the different data sets correspond to
    simulations with different injection rates $Q$, and thus to different
    capillary numbers Ca. The values of Ca are indicated in the figure legends.
    The average cumulative growth density $\Phi(z)$ is plotted as solid black
    line, which shows the agreement with Eq. \ref{eqn_sigmac}. The black, blue
    and red dashed lines are plotted using Eq. \ref{eqn_saffman} with
    $\lambda=0.4$, $0.5$ and $0.62$, which correspond to the experimental
    observation for porous Hele-Shaw cell \cite{lmt04,tlm05}, the DLA
    \cite{aet96,sb03} and the Saffman-Taylor solution for low surface tension
    limit \cite{st58} respectively. The insets show the variations of unscaled
    $\sigma(z)$ for different values of Ca.}
\end{figure*}

\begin{figure}[htbp]
    \captionsetup[subfigure]{labelformat=empty}
    \centerline{
    \subfloat[$\mathrm{Ca}=0.04$]{\includegraphics[width=0.14\textwidth]{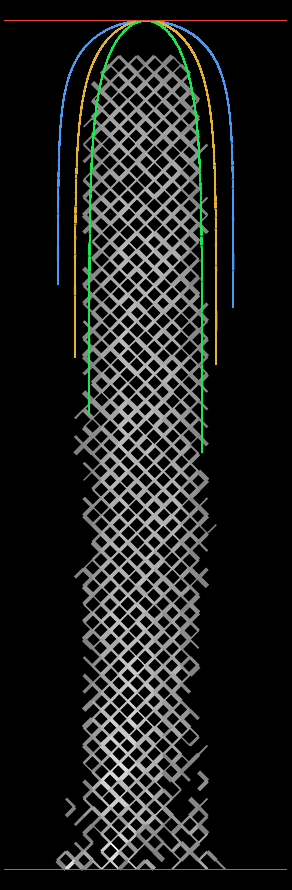}}\hfill
    \subfloat[$\mathrm{Ca}=0.08$]{\includegraphics[width=0.14\textwidth]{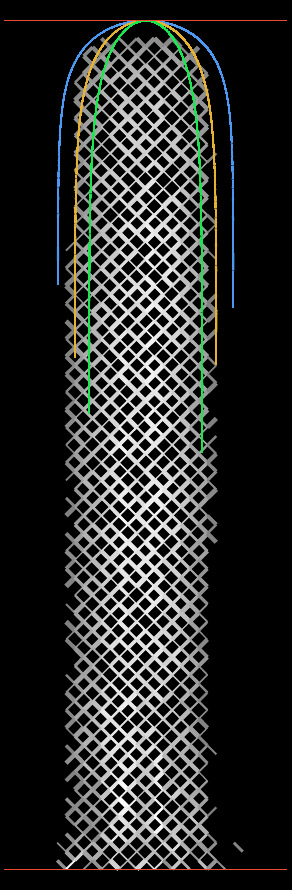}}\hfill
    \subfloat[$\mathrm{Ca}=0.1$]{\includegraphics[width=0.14\textwidth]{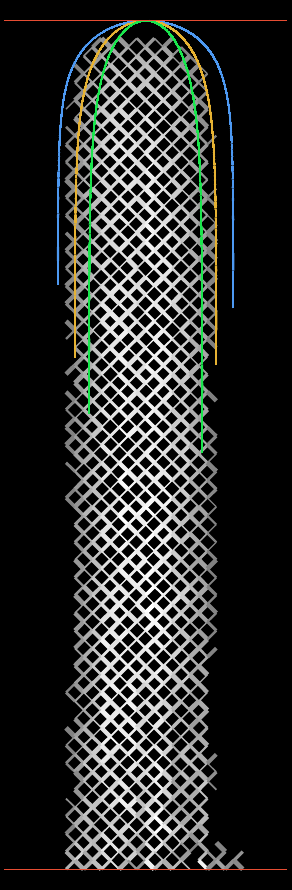}}}
    \caption{\label{fig_threshold}Volume density maps thesholded with
    $\tilde\sigma(x,z) \ge\tilde\sigma_\text{max}/2$ for $M=10^{-4}$ and for
    $\text{Ca}=0.04$, $0.08$, and $0.1$, as indicated below the figures. The
    gray shades correspond to the same density map as shown in Fig.
    \ref{fig_snapshots} for the respective values of Ca, but now only the links
    preserved by the cut-off are shown. The green, yellow and blue lines are
    drawn using the Saffman-Taylor equation (Eq. \ref{eqn_saffman}) with
    $\lambda = 0.4$, $0.5$ and $0.62$, which correspond to the experimental
    observation for porous Hele-Shaw cell, the continuum-medium Saffman-Taylor
    finger and the DLA finger respectively.}
\end{figure}

Note that $\sigma(z)$ and $\phi(z)$ are related to each other. Indeed, let us
now examine the displacement process in the referential $(x,\tilde{z})$ attached
to the inlet boundary of the network, with $\tilde{z}$ and $z$ related to each
other according to $z=\tilde{z}_{\rm tip} (t) - \tilde{z}=v_{\rm tip} t  -
\tilde{z}$, where $v_{\rm tip}$ is the velocity of the most advanced tip of the
non-wetting fluid phase and $\tilde{z}_{\rm tip} (t)$ is its position along
$\tilde{z}$. That velocity is constant in time, as shown in Fig.
\ref{fig_tipvel}, where $\tilde{z}_{\rm tip}$ is plotted as a function of the
time normalize by the time $t$ scaled by $t_f$, where $t_f$ is the time at which
the tip reaches the outlet boundary of the porous network. At any time $t$, the
volume of non-wetting fluid in a slice of the network normal to the mean flow
direction and located between $\tilde{z}$ and $\tilde{z}+\Delta \tilde{z}$ is
$\tilde{\sigma}(\tilde{z},t)\Delta \tilde{z}$, with $\Delta \tilde{z}=\Delta z$.
It results from the cumulative growth of the fingers between the time
$t_0=\tilde{z}/v_{\rm tip}$ at which the most advanced tip first arrived at
$\tilde{z}$ and $t$, and can thus be expressed as 
\begin{equation}
    \tilde{\sigma}(\tilde{z},t) \Delta z = \Delta z  \int_{\tilde{z}/v_{\rm tip}}^t \tilde{\nu}(\tilde{z},u)\dif u ~ .
\end{equation}
The steady volumetric density defined above in the $(x,z)$ reference frame,
$\sigma(z)$, is related to $\tilde{\sigma}(\tilde{z},t)$ through the following
relation:
\begin{equation}
    \forall t ~~~ \sigma(z)= \tilde{\sigma} \left (v_{\rm tip} t - z,t \right )~.
\end{equation}
A similar expression relates the volumetric growth rate expressed in the moving
reference frame, $\nu(z)$, and that expressed in the laboratory frame,
$\tilde{\nu}(\tilde{z},t)$. Hence,
\begin{equation}
    \label{eqn_intphi_new}
    \begin{aligned}
        \displaystyle
        \forall t  ~~ ~    \sigma(z) \Delta z & =  \Delta z  \int_{t_0}^t \tilde{\nu} \left  (v_{\rm tip} t' - z,t' \right )\dif t' ~ .\\
        \mathrm{i.e.,~ ~}    \sigma(z) \Delta z   & =  \Delta z  \int_0^z  \nu(z') \frac{\dif z'}{v_{\rm tip}} \\
        \mathrm{i.e.,~ ~}    \sigma(z) \Delta z & = \frac{Q\Delta z}{v_{\rm tip}} \int_0^z  \phi(z') \dif z' ~.
    \end{aligned}
\end{equation}
Defining the cumulative growth density in the $z$ direction as $\Phi(z)=\int_0^z
\phi(z')\dif z'$ and $\sigma_{\rm c}=Q/v_\text{tip}$ leads to
\begin{equation}
  \label{eqn_sigmac}
  \displaystyle
  \Phi(z) = \frac{\sigma(z)}{\sigma_{\rm c}}~,
\end{equation}
which relates the volume density with the cumulative growth density.

\begin{figure*}[htbp]
    \centerline{ \includegraphics[width=0.32\textwidth]{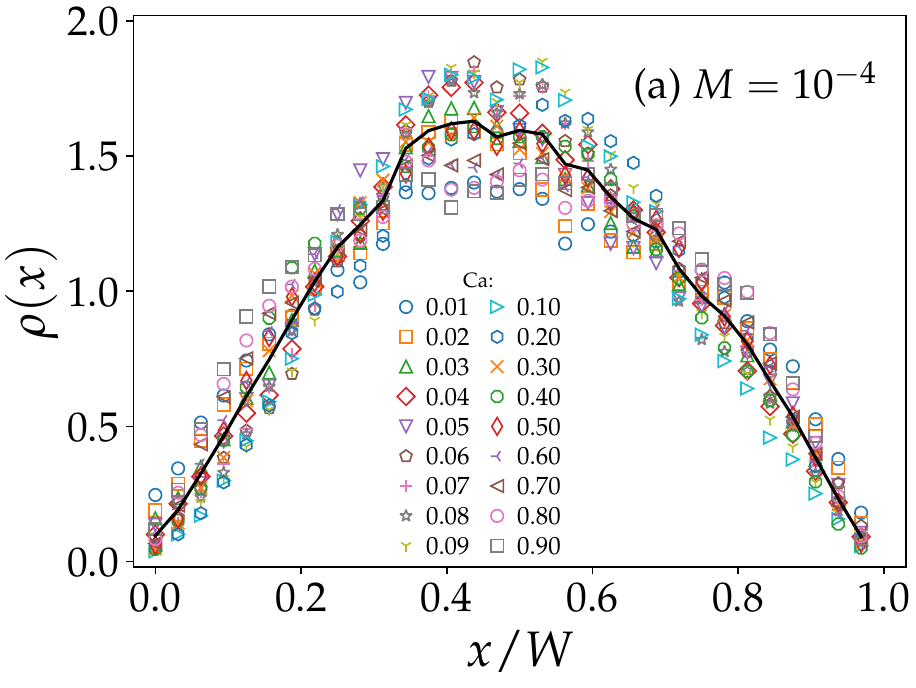}\hfill
        \includegraphics[width=0.32\textwidth]{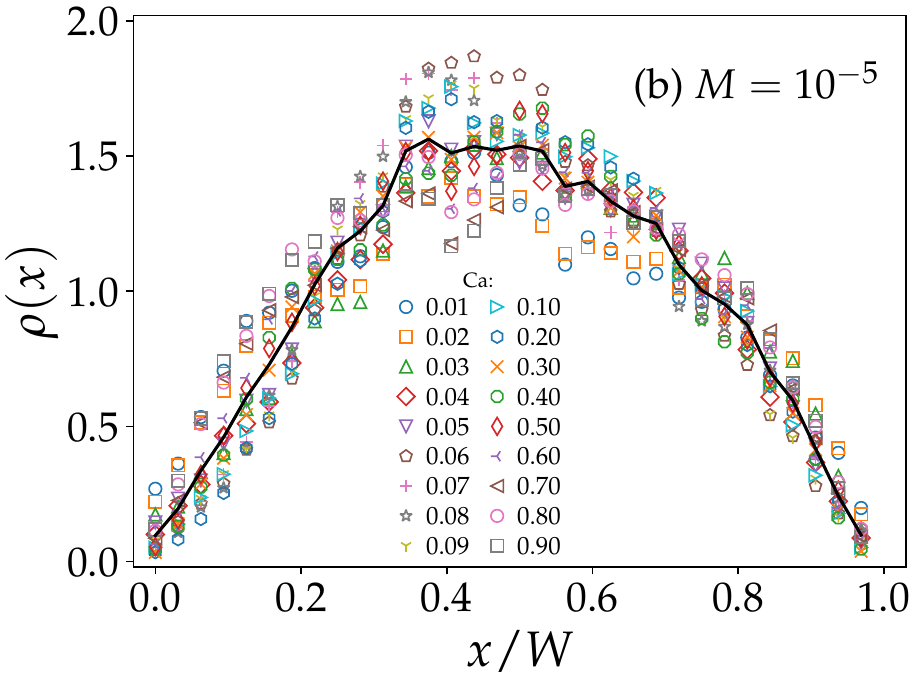}\hfill
        \includegraphics[width=0.32\textwidth]{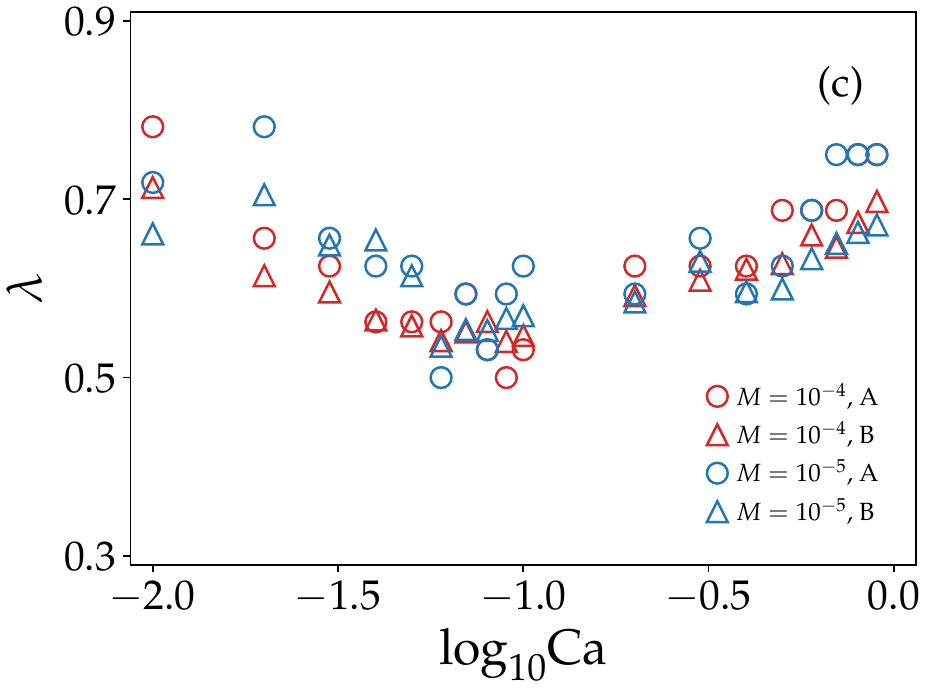}}
        \caption{\label{fig_lambda} Plots of the average transverse volume
        density profile $\rho(x)$ as a function of the transverse normalized
        coordinate $x/W$ are shown for (a) $M=10^{-4}$ and (b) $M=10^{-5}$. The
        different symbols show the results for different values of Ca and the
        solid line shows the profile averaged over all datasets. The values of
        $\lambda$ measured using the two different methods, (A)
        $\lambda=1/\rho_\text{max}$ and (B) $\lambda = (x_+-x_-)/W$ where
        $\rho(x_+)=\rho(x_-) = \lambda_\text{max}/2$, are shown in (c) as a
        function of the capillary number.}
\end{figure*}

To verify this functional form, the tip velocity is first measured from the
slopes as  the straight lines in figure \ref{fig_tipvel} multiplied by $t_f$. We
then plot $\sigma(z)$ scaled by $\sigma_c$, for different values of Ca and $M$
(Fig. \ref{fig_volden}). Notice that, though $\sigma(z)$ is a function of the
capillary numbers as shown in the insets, all $\sigma(z)/\sigma_c$ plots
collapse on a single master curve for different values of Ca. With Eq.
\ref{eqn_sigmac}, it therefore implies that there is one unique cumulative
growth density $\Phi(z)$, independent of Ca. We therefore calculate the average
cumulative growth density $\Phi(z)$ from the data in Fig. \ref{fig_growthden}
and plot it in Fig. \ref{fig_volden} as a solid black line. The line shows an
excellent agreement with the $\sigma(z)/\sigma_c$ plots, as expected from Eq.
\ref{eqn_sigmac}. Such a growth property of the front is the characteristics of
the viscous dominated regime, or of a regime where viscous forces compete with
capillary forces, and therefore will not be observed in pure capillary fingering
regime where the invasion is completely controlled by the disorder in the
capillary thresholds.

\subsection{The `continuum' shape}
\label{sec_shape}
The average volume densities of the fractal fingers can be used to compare their
shapes statistically with the smooth shapes of continuum-medium fingers, as well
as with the shapes of DLA. By mapping the volume density function $\sigma(z)$ to
the Saffman Taylor equation (Eq. \ref{eqn_saffman}) for continuum viscous
fingers, one can calculate the width ratio $\lambda$, and use that to compare
the statistical shape profile of different fingers. Before presenting the
detailed measurements of $\lambda$, we show a quick comparison in Fig.
\ref{fig_threshold} where only the part of the network with volume density
$\tilde\sigma(x,z)\ge\sigma_\text{max}/2$ are shown. This region corresponds to
the porous media with large invasion \cite{caa90,aac91,aet96} and is comparable
to the Saffman-Taylor equation. For the viscous fingers in continuum Hele-Shaw
cell without any porous structure, the width ratio $\lambda$ in Eq.
\ref{eqn_saffman} has a value of $0.5$ in the limit of negligible surface
tension, and it increases with the increase of surface tension
\cite{ms81,cdh86}. For experiments with porous Hele-Shaw cells on the other
hand, $\lambda$ was found to be around $0.4$ \cite{lmt04,tlm05}, whereas for
off-lattice DLAs, $\lambda$ was found to be around $0.62$ for linear channels
\cite{aet96,sb03}. The solutions of Eq. \ref{eqn_saffman} for these three values
are shown in Fig. \ref{fig_threshold}, where the green, yellow and blue lines
correspond to $\lambda=0.4$, $0.5$ and $0.62$ respectively. The lines show
deviations from the gray-scale region and the deviations also depend on Ca. We
also show the solutions of Eq. \ref{eqn_saffman} for these three values of
$\lambda$ in Fig. \ref{fig_volden} by the dashed lines, which again show some
deviations from the results for the drainage fingers here.

In order to measure the value of the width parameter $\lambda$, we measure a
transverse volume density $\rho(x)$ in a region where the growth is almost
frozen, that is, far behind the most advanced fingertip. If $L_{\rm f}$ is the
length of this frozen zone in the $z$ direction, then one has
\begin{equation}
    \label{eqn_frozensum}
    \displaystyle
    \int_{0}^{W} \int_{W}^{W+L_{\rm f}}\tilde\phi(x)\dif z\dif x = \sigma_cL_{\rm f} \;,
\end{equation}
as $L_{\rm f}$ increases with the same velocity $v_\text{tip}$. We therefore
define $\rho(x)$ as,
\begin{equation}
  \label{eqn_voldenx}
  \displaystyle
  \rho(x) = \frac{W}{\sigma_cL_{\rm f}}\int_{W}^{W+L_{\rm f}}\tilde\phi(x,z)\dif z \;,
\end{equation}
with $\int_0^W\rho(x)\dif x=W$. In Fig. \ref{fig_lambda}a and \ref{fig_lambda}b,
we plot $\rho(x)$ as a function of the scaled transverse position $x/W$ for the
two different viscosity ratios. The datasets show a maximum at the middle and
then decrease on both sides. This is an expected behavior, similar to what is
observed for DLA \cite{aet96} and for the experiments of viscous fingers in
porous media \cite{lmt04,tlm05}. However, the plots drift with Ca, indicating a
variation of $\lambda$ with Ca. From this data, the width ratio $\lambda$ can be
estimated in two ways: either from (A) $\lambda = (x_+-x_-)/W$ where
$\rho(x_+)=\rho(x_-) = \lambda_\text{max}/2$ or alternatively from (B)
$\lambda=1/\rho_\text{max}$ \cite{caa90,aac91}. The two measurements are shown
in Fig. \ref{fig_lambda}c, where $\lambda$ seems to vary systematically with Ca.
Interestingly, the lowest value here is at $\approx 0.5$, which is the solution
for the continuum Saffman-Taylor fingers for negligible surface tension limit
\cite{st58}. Higher values of $\lambda$ were found for increasing surface
tension for the continuum fingers \cite{cdh86,ms81,v83}. The value of
$\lambda\approx 0.62$ for DLAs in linear channels \cite{aet96,sb03} coincides
with some of the data points here, however the two-phase flow experiments in
porous Hele-Shaw cells showed a much lower value of $\lambda\approx 0.4$
\cite{lmt04,tlm05}. Note however that in contrast to the present system, the
capillary threshold distribution was not uniform in these experiments, despite
the fact that a uniform distribution was assumed for some of the
interpretations. Here for our system, $\lambda$ increases up to a much higher
value of around $0.8$. Such values are similar to the fingers in divergent
cells, which are linear cells with an angular wedge at the inlet, a geometry
that is in between the linear and circular geometries \cite{trh89}. There,
$\lambda$ was found to increase when increasing the angle of the wedge, for
example $0.77$ to $0.82$ for a wedge angle of $90^\circ$ \cite{aet96,acg89}.
Note that, in our simulations, we inject fluid only through a few of the nodes
at the center of the network's inlet boundary, and all other nodes on the two
sides of these injection nodes are blocked, see Fig. \ref{fig_network}. This may
have a wedge effect on the value of $\lambda$, similar to the divergent inlets.
However, the fact that here $\lambda$ shows a minimum at an intermediate Ca and
increases on both sides, indicates that there is a combined effect of surface
tension and inlet geometry, and it therefore needs further in-depth study. We
leave this for future, as our main focus here is to explore the effect of
disorder on the relationship between the growth and pressure drop.

\subsection{Growth rate versus local pressure drop}
\label{sec_pressure}
The volumetric growth of the region occupied by the invading fluid inside a pore
depends on whether the viscous pressure drop between the invading and defending
fluids is able to overcome the capillary barrier inside the pore and thus to
displace the interface between the two fluids. Assuming that the movement of the
interfaces follow Eq. \ref{eqn_washburn}, a theoretical methodology was
suggested in \cite{lmt04,tlm05} to find the relationship between the growth rate
and the local pressure drop across the fluid-fluid interface by integrating over
the distribution of the capillary barriers. We follow their approach in the
following and modify it to adapt our system. If $P'$ is the threshold pressure
required to invade a link and $g(P')$ is the normalized distribution of the
thresholds over the network, we may integrate over $g(P')$ to obtain the growth
rate,
\begin{equation}
    \label{eqn_intq}
    \displaystyle
    \begin{aligned}
        \tilde\nu(x,z) & = \frac{1}{h^2}\int q(x,z) g(P')\dif P' \\
        = \frac{\kappa}{h^2} &\int (\Delta p(x,z) - P')\Theta(\Delta p(x,z) - P') g(P')\dif P' \;,
    \end{aligned}
\end{equation}
where $\Delta p(x,z)$ is the pressure drop between the invading and defending
fluids inside a link $i$ at ($x,z$) and $q(x,z)$ is the corresponding flow rate
of the fluids in that link. Here we consider $\kappa=\bar k/\mu_w$, where $\bar
k$ is the average mobility of the pores and $\mu_w$ is the viscosity of the
high-viscosity defending fluid. A pore will be invaded only when $\Delta
p(x,z)>P'$, hence the Heaviside function $\Theta(\Delta p(x,z)-P')$.

\begin{figure}
    \centerline{\hfill\includegraphics[width=0.45\textwidth]{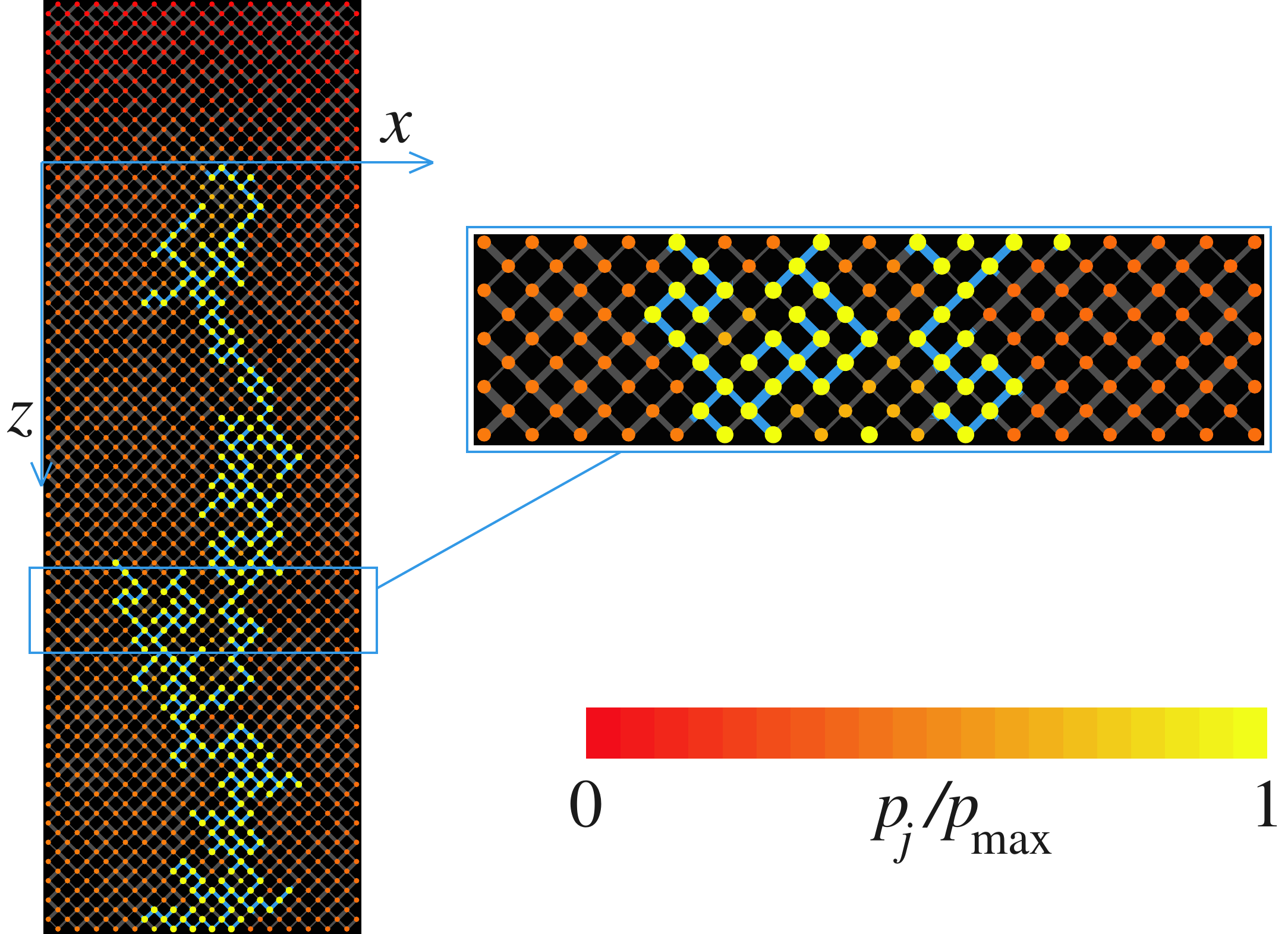}\hfill}
    \caption{\label{fig_presdemo} Node pressures ($p_j$) obtained from conjugate
    gradient solver, normalized by the maximum pressure $p_{\rm max}$ at that
    time step, shown by colors for a network of $32\times 96$ links. The color
    scale shows the normalized pressure values, where the darkest red
    corresponds to $p_j=0$ and the lightest yellow correspond to the highest
    pressure. Here $\text{Ca}=0.05$ and $M=10^{-5}$. The injected fluid is
    colored blue. The box at the top right shows a close view of a small slice
    of the network. Notice that the nodes belonging to the low-viscosity finger
    are at much higher pressures than those belonging to the defending fluid.}
\end{figure}

If we consider $P'$ to be uniformly distributed between $P_{\rm t}$ and $P_{\rm
m}$, in the form
\begin{equation}
    \label{eqPtDist}
    \displaystyle
    g(P') = \frac{1}{G}\Theta(P'-P_{\rm t})\Theta(P_{\rm m}-P')
\end{equation}
where $G=P_{\rm m}-P_{\rm t}$, then Eq. \ref{eqn_intq} becomes
\begin{equation}
    \label{eqn_intq2}
    \displaystyle
    \tilde\nu(x,z)  = \frac{\kappa}{h^2 G} \int_{P_{\rm t}}^{P_{\rm m}} (\Delta p - P')\Theta(\Delta p - P')  \dif P' \;,
\end{equation} 
where $\Delta p \equiv \Delta p(x,z)$. If the viscous pressure drop exceeds the
capillary threshold for all pores, i.e., if $\Delta p(x,z)>P_{\rm m}$, then Eq.
\ref{eqn_intq2} reduces to
\begin{equation} 
    \label{eqn_intq3}
    \displaystyle
    \begin{aligned}
        \tilde\nu(x,z) &  = \frac{\kappa}{h^2 G} \int_{P_{\rm t}}^{P_{\rm m}} (\Delta p - P') \dif P' \\
        &  =  \frac{\kappa}{2 h^2 (P_{\rm m}-P_{\rm t})}  \left [ - \left ( \Delta p-P'\right ) ^2 \right ]_{P_{\rm t}}^{P_{\rm m}} \\
        \tilde\nu(x,z) &  = \frac{\kappa}{h^2} \left [ \Delta p -   \frac{P_{\rm m}+P_{\rm t}}{2} \right ]\;.
    \end{aligned}
\end{equation}
The growth rate in this regime then varies linearly with the excess pressure
drop with respect to the mean capillary threshold $(P_{\rm m}+P_{\rm t})/2$.

For $\Delta p(x,z)<P_{\rm t}$ on the other hand, Eq. \ref{eqn_intq2} reduces to
\begin{equation} 
    \label{eqn_intq4}
    \displaystyle
    \begin{aligned}
        \tilde\nu(x,z) &  = \frac{\kappa}{h^2 G} \int_{P_{\rm t}}^{\Delta p} (\Delta p - P') \dif P' \\
        \tilde\nu(x,z) &  = \frac{\kappa}{2 h^2} \frac{\left ( \Delta p -   P_{\rm t} \right ) ^ 2}{P_{\rm m}-P_{\rm t}} \propto \left ( \Delta p -   P_{\rm t} \right ) ^ 2 \;.
    \end{aligned}
\end{equation}
The growth rate thus varies quadratically with $[\Delta p(x,z)-P{\rm t}]$. This
regime corresponds to moderate capillary numbers where the pressure drop $\Delta
p(x,z)$ competes with the capillary thresholds, and the number of invaded pores
thus  increases with an increase in $\Delta p(x,z)$.

If we further assume that $\Delta p(x,z)$ is a function of $z$ only, we may
average $\Delta p(x,z)$ over $x$, which yields
\begin{equation}
    \label{eqn_delpquadz}
    \displaystyle
    \nu(z) \sim \left ( \Delta P(z)-P_{\rm t}\right )^2  \;,
\end{equation}
where $\Delta P(z)=\langle\Delta p(x,z)\rangle_x$. Hence, in this regime, the
average growth rate $\nu(z)$ depends quadratically on the excess pressure drop
at $z$, $[\Delta P(z)-P_{\rm t}]$. Such a quadratic relationship between excess
pressure drop and flow rate was also suggested for steady state two-phase flow
in porous media by mean-field calculations \cite{sh12}, and observed by
numerical simulations \cite{sh12,sbd17,yts13}.

\begin{figure*}[htbp]
    \centerline{\hfill
    \includegraphics[width=0.32\textwidth]{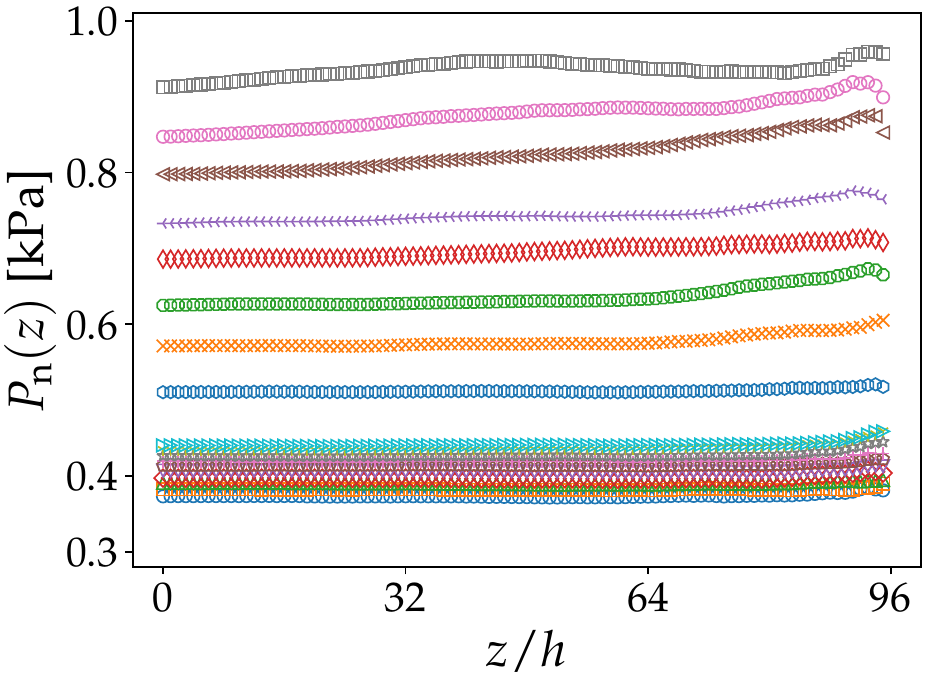}\hfill
    \includegraphics[width=0.32\textwidth]{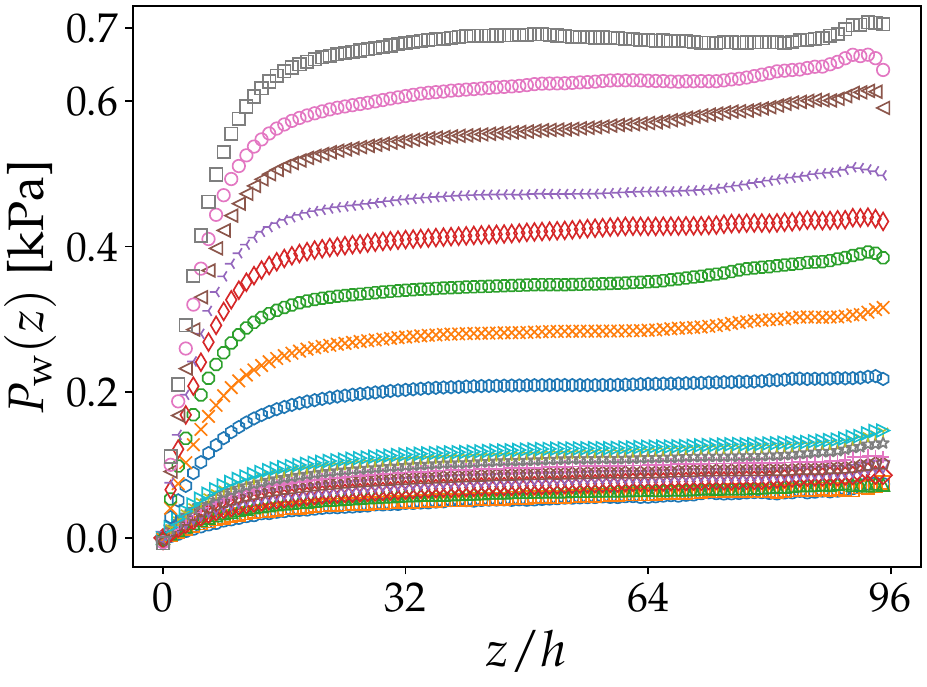}\hfill
    \includegraphics[width=0.32\textwidth]{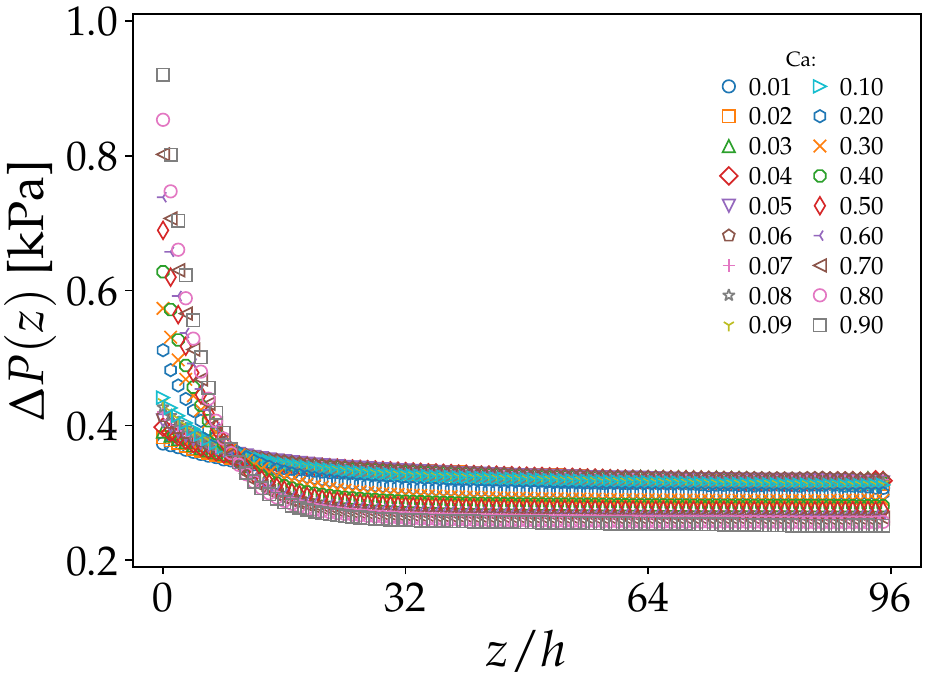}\hfill}
    \centerline{\hfill
    \includegraphics[width=0.32\textwidth]{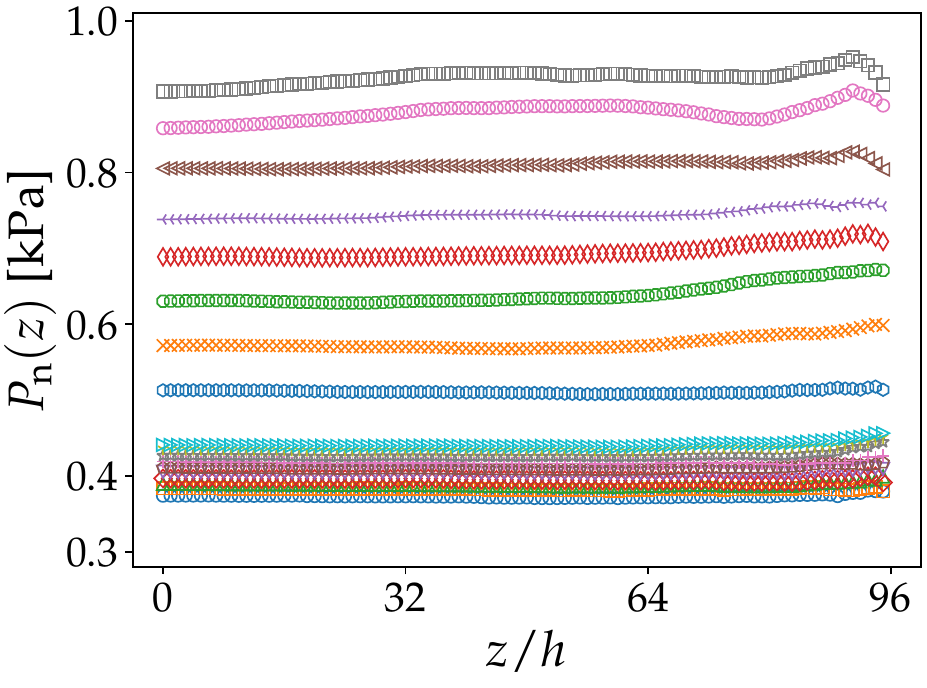}\hfill
    \includegraphics[width=0.32\textwidth]{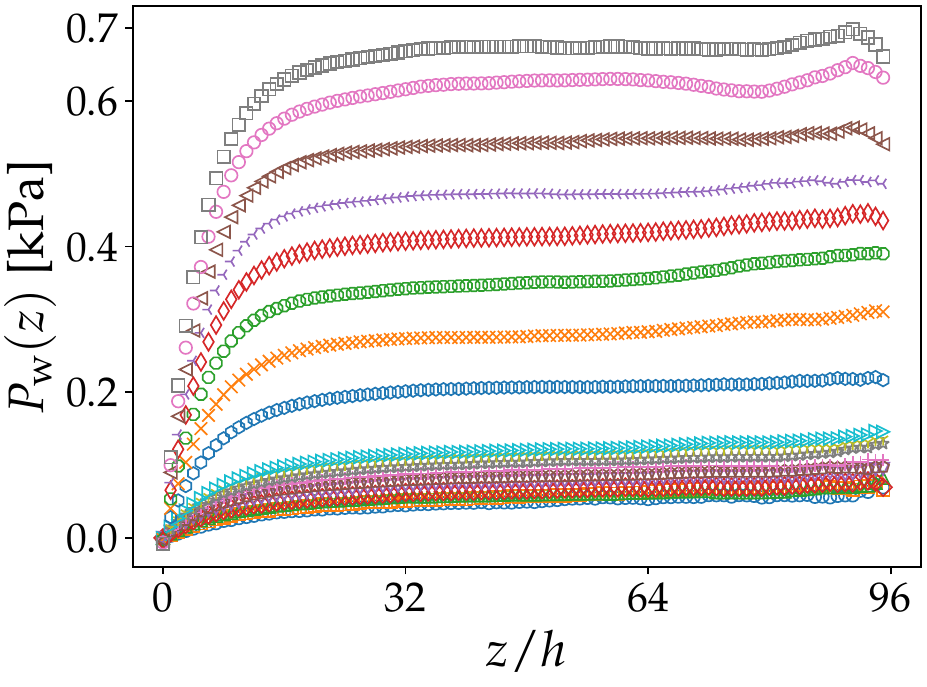}\hfill
    \includegraphics[width=0.32\textwidth]{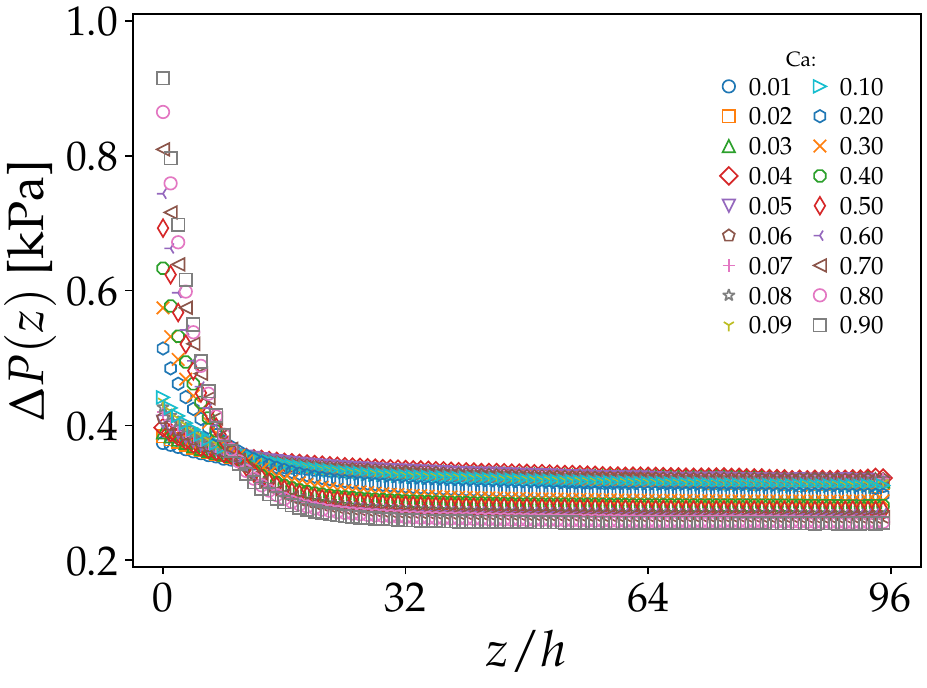}\hfill}
    \caption{\label{fig_pressure} Plots of the non-wetting pressure $P_{\rm
    n}(z)$, the non-wetting pressures $P_{\rm w}(z)$ and the pressure drop
    $\Delta P(z)$, obtained using Eq. \ref{eqn_presmeasure} and \ref{eqn_pdiff},
    as a function of the distance from the most advanced fingertip $z/h$, for
    different capillary numbers. The top row corresponds to $M=10^{-4}$ and the
    bottom row to $M=10^{-5}$.}
\end{figure*}

Note however that the threshold distribution function $g(P_{\rm t})$ considered
in the above derivation was assumed to be uniform. In our pore network it is  a
more complex function, which is not straightforward to determine. It not only
depends on the pore properties such as the size distribution \cite{rsh21} and
the pore wettabilities \cite{fsr21,fsh23} but also on the structure of the
finger itself, that is, the distribution of the number of fingertips at any $z$.
We therefore may assume a generalization of Eq. \ref{eqn_delpquadz},
\begin{equation}
    \label{eqn_delpbetaz}
    \displaystyle
    \nu(z) \sim \left ( \Delta P(z)-P_{\rm t}\right )^\beta \;,
\end{equation}
where $\beta$ is an exponent of non-linearity. As mentioned in the Introduction,
such non-linear relationships with $\beta > 1$ have been widely observed for
steady-state flow in a certain regime of capillary numbers \cite{tkr09, tlk09,
aet14, rcs11, rcs14, sbd17, glb20, zbg21, zbb22, sh12, shb13, rhs19, cfh23,
sh12, yts13}.

In order to examine the form of Eq. \ref{eqn_delpbetaz}, we need to measure two
quantities: (a) the average pressure difference between the defending and
invading fluids, $\Delta P(z)$, as a function of $z$, and (b) the threshold
pressure $P_{\rm t}$. At every time step, we obtain the pressures at the nodes
of the network from the conjugate gradient solver by solving the equations
\ref{eqn_washburn} and \ref{eqn_young}. An example is shown in Fig.
\ref{fig_presdemo}, where the pressure values are indicated by different colors.
Notice that the nodes connected by the invading non-wetting low-viscosity fluid
(blue links) are at much higher pressures than those that are connected with the
defending wetting fluid. The conjugate-gradient solver does not distinguish
between the wetting and non-wetting pressures; we therefore identify, at every
time step, the links filled with invading fluid as those with a non-wetting
saturation larger than a threshold value ($s_i>0.98$), and the nodes that are
connected to them are marked as the non-wetting nodes. The rest of the nodes are
then marked as the wetting nodes. We then measure the average excess wetting and
non-wetting pressures as a function of the distance $z$ from the most advanced
fingertip as 
\begin{equation}
    \label{eqn_presmeasure}
    \begin{aligned}
        \displaystyle
        P_{\rm w}(z) & = \frac{1}{n_{\rm w}(z)}\sum_{k=1}^{n_{\rm w}(z)} p_k(z) - P_0 \\
        \text{and~ ~}   P_{\rm n}(z) & = \frac{1}{n_{\rm n}(z)}\sum_{k=1}^{n_{\rm n}(z)} p_k(z) - P_0 \; ,
    \end{aligned}
\end{equation}
where $n_{\rm w}(z)$ and $n_{\rm n}(z)$ are the number of wetting and
non-wetting nodes at $z$, and $P_0$ is the average wetting pressure at the most
advanced fingertip, $z=0$. The pressure drop $\Delta P(z)$ between the wetting
and non-wetting fluids at $z$ is then calculated as
\begin{equation}
  \label{eqn_pdiff}
  \displaystyle
  \Delta P(z) = P_{\rm n}(z) - P_{\rm w}(z) \;,
\end{equation}
which is then averaged over different time steps during the propagation of the
fingers, and over different samples.

\begin{figure*}[htbp]
    \centerline{\hfill
    \includegraphics[width=0.45\textwidth]{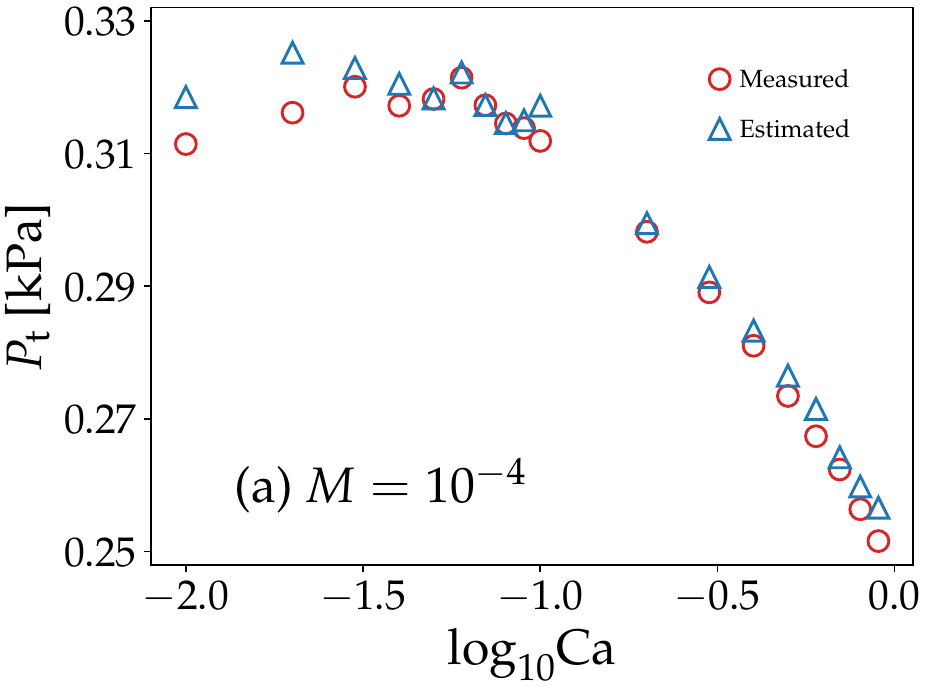}\hfill
    \includegraphics[width=0.45\textwidth]{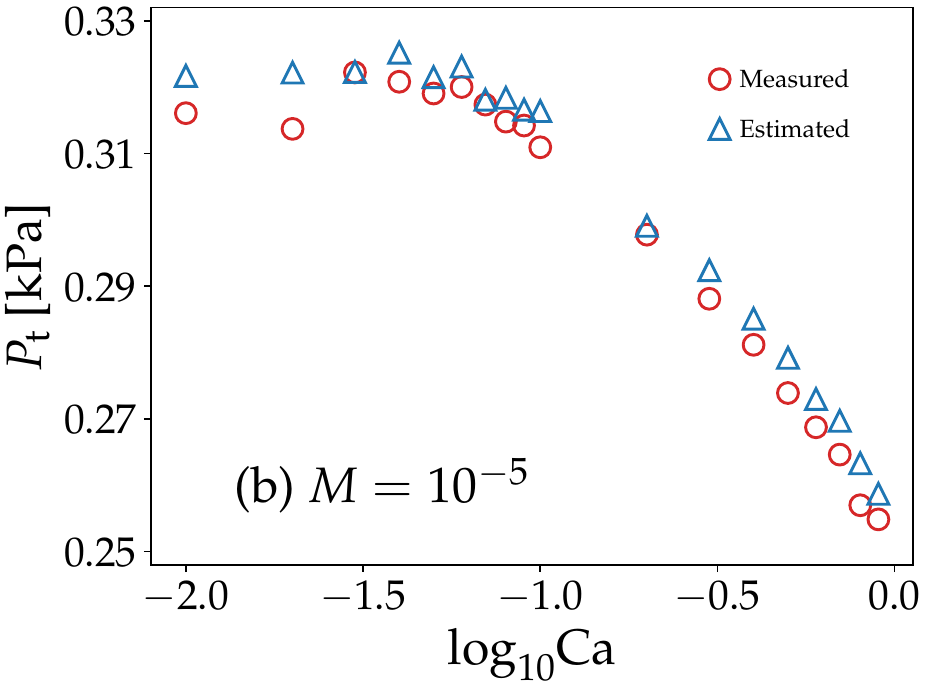}\hfill}
    \caption{\label{fig_pt} Dependence of the threshold pressures $P_{\rm t}$ on
    the capillary numbers Ca for the two different viscosity ratios ($M$), (a)
    $10^{-4}$ and (b) $10^{-5}$. The {\it measured} values of $P_{\rm t}$
    represented by the red circles are obtained by taking the averages of
    $\Delta P(z)$ for $z/h>76$, i.e. over the last $20\%$ of the data points on
    the plateau of $\Delta P(z)$ in Fig. \ref{fig_pressure}. The {\it estimated}
    values of $P_{\rm t}$ represented by the blue triangles are obtained by
    minimizing the least-square fit error of Eq. \ref{eqn_delpbetaz}.} 
\end{figure*}

In Fig. \ref{fig_pressure}, $P_{\rm n}(z)$, $P_{\rm w}(z)$ and $\Delta P(z)$ are
plotted as functions of the row distance $z/h$ from the fingertip for different
values of Ca and $M$. We can notice a few things here. First, the wetting
pressure $P_{\rm w}$ shows a sharp variation with $z$, whereas the non-wetting
pressure $P_{\rm n}$ do not show any noticeable variation. This is because of
the low viscosity of the invading fluid compared to the defending fluid.
Secondly, the pressure drop $\Delta P(z)$ starts from a high value at $z=0$,
then falls rapidly withing a distance $z<W/2$ and almost goes to a plateau. This
behavior is similar to the growth density $\phi(z)$ in Fig. \ref{fig_growthden},
which showed a rapid exponential fall within a similar distance, indicating that
the growth $\phi(z)$ is directly related to $\Delta P(z)$. Furthermore, note
that the plateau regimes of $\Delta P$ far behind the tip has finite values
instead of zero. This however is the stagnant zone, where the fingers do not
show any visible growth and $\phi(z)$ is essentially zero as seen in Fig.
\ref{fig_growthden}. Therefore in this regime, the capillary barriers compete
with the viscous pressure drop and prevent the interfaces from moving. The
plateau value thus corresponds to the threshold pressure $P_{\rm t}$ in Eq.
\ref{eqn_delpbetaz}, below which there is no growth. We can therefore measure
$P_{\rm t}$ from this plateau regime, by taking the average of $\Delta P(z)$
over its farthest data points. However, if we look carefully on the plots, we
notice that not all data sets at the plateau exhibit a perfectly constant
behavior, some of them are still decreasing slowly and therefore would need a
much larger system to saturate. We therefore estimate the threshold pressure
$P_{\rm t}$ in one more way where we minimize the least-square fit error for Eq.
\ref{eqn_delpbetaz}. We try a set of trial values for $P_{\rm t}$ around the
plateau of $\Delta P(z)$ and then choose the one that provides the lowest
least-square error for the straight-line fit for $\log[\nu(z)]$ vs $\log[\Delta
P(z)-P_{\rm t}]$. These {\it estimated} values of $P_{\rm t}$ are shown in Fig.
\ref{fig_pt} with blue triangles, where we also compare them with the {\it
measured} values of $P_{\rm t}$, which are calculated as the average of $\Delta
P$ over the farthest $20\%$ rows from the fingertip. Notice that the values of
$P_{\rm t}$ from these two different calculations are very close to each other
for the whole range of the capillary numbers. Furthermore, $P_{\rm t}$ does not
seem to depend on the viscosity ratio $M$, which confirms that the threshold is
a quantity controlled solely by the capillary barriers, which are functions of
the interfacial tension between the two fluids.

\begin{figure*}[htbp]
    \centerline{\hfill
    \includegraphics[width=0.45\textwidth]{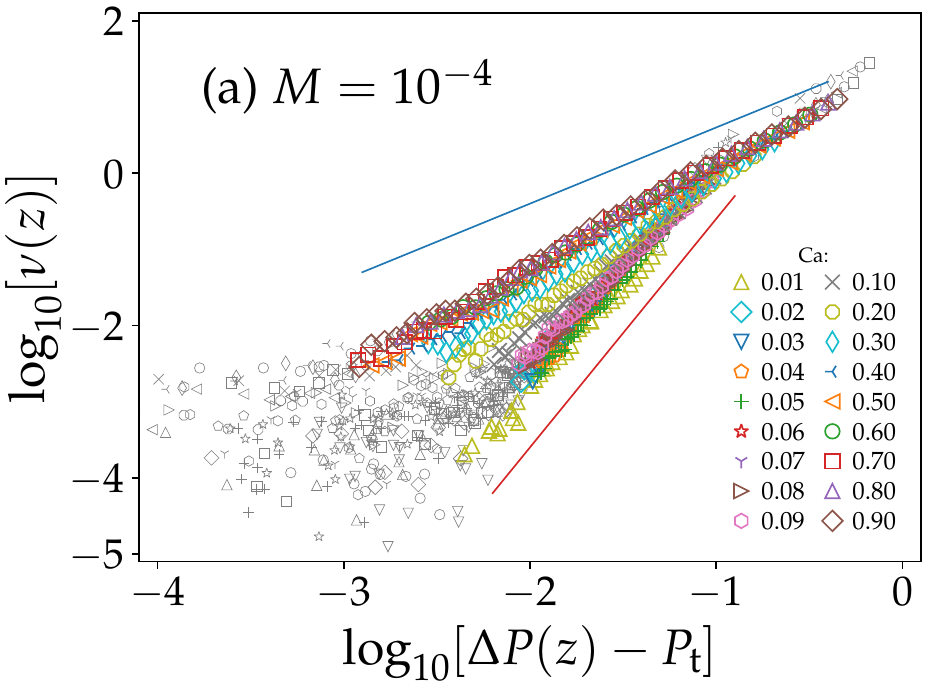}\hfill
    \includegraphics[width=0.45\textwidth]{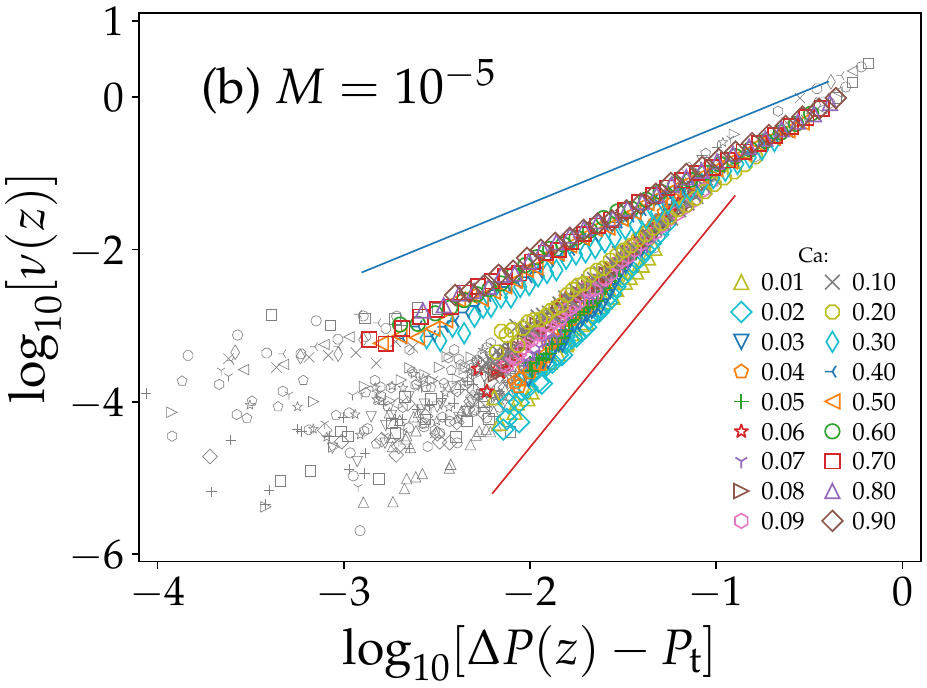}\hfill}
    \caption{\label{fig_logfit}Plot of the local excess pressure drop
    $\log_{10}[\Delta(P(z)-P_{\rm t}]$ between the invading and defending fluids
    as a function of the local growth rate of the finger, $\log_{10}[\nu(z)]$,
    for the viscosity ratios (a) $M=10^{-4}$ and (b) $M=10^{-5}$. Different sets
    correspond to different values of Ca as indicated by the symbols. The
    exponents $\beta$ defined in Eq. \ref{eqn_delpbetaz} is calculated from the
    slopes. The slopes are calculated only for the range plotted with colored
    symbols, as the growth far behind the fingertip is almost frozen and lead to
    noisy data points as shown by the gray symbols. The blue and red straight
    lines are drawn with slopes $1$ and $3$.} 
\end{figure*}

\begin{figure}[htbp]
    \centerline{\hfill\includegraphics[width=0.45\textwidth]{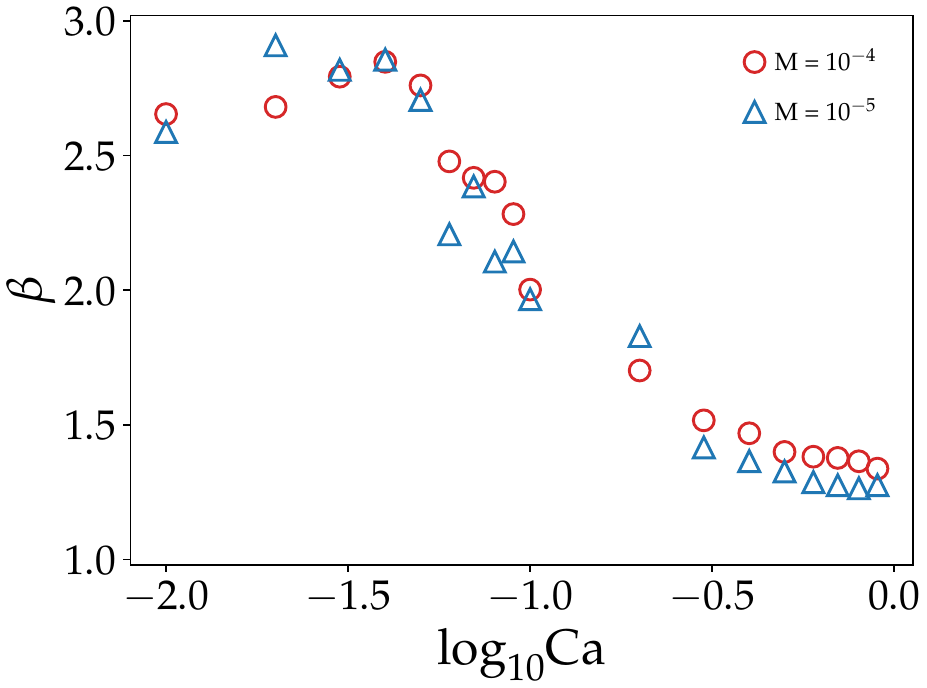}\hfill}
    \caption{\label{fig_beta}Dependence of the growth exponent $\beta$ obtained
    from Fig. \ref{fig_logfit} on Ca. The error bars are of order $0.07$ towards
    the lower side of Ca whereas they are $0.01$ on the higher side. They are
    too small compared to the scale of the plot, and therefore not shown. The
    two types of symbols correspond to the two values of $M$ as indicated in the
    legend of the plot.} 
\end{figure}

Finally, we now set out to verify the agreement with Eq. \ref{eqn_delpbetaz} and
determine the exponent $\beta$. In Fig. \ref{fig_logfit}, we plot the growth
rate $\nu(z)$ as a function of the excess local pressure drop $[\Delta
P(z)-P_{\rm t}]$ in $\log$ scale, where we use the values of $P_{\rm t}$
estimated from Fig. \ref{fig_pt}. The plots show a linear trend over one to
three decades, depending on the capillary number, for the entire range of
capillary numbers, with a linear range of the log-log plot that is all the
larger as Ca is larger. The power law exponents vary systematically with Ca
within a range of $1$ to $3$ as shown by the blue and red straight lines in the
plots. However, at very small values of the pressure drops, the data is noisy
and deviates from the linearity, which corresponds to the gray symbols in the
plots. This corresponds to the zone far behind the fingertip, where the $\Delta
P(z)$ is very close to $P_{\rm t}$ and the growth structure is almost frozen. As
mentioned before, the statistical averaging becomes poorer as we move away from
the fingertip, which therefore adds to the statistical noise in that region. We
have therefore disregarded these points when fitting power laws to the data, and
only rely on the points shown as colored symbols in the plots. The exponent
$\beta$ is thus measured from least-square fitting a power law to the colored
data points. In figure \ref{fig_beta} we plot the dependence of $\beta$ on Ca
for the two viscosity ratios, which exhibit very similar trend, i.e., two
plateaus for $\mathrm{Ca} \lesssim 10^{-1.3}$ and $\mathrm{Ca} \gtrsim
10^{-0.2}$, and a crossover in between them. On the lower capillary number
plateau, the values fluctuate within a range of $\approx 2.5$ to $3.0$. It then
undergoes a crossover with the increase of Ca, and then approaches another
plateau towards $1$. This is very similar to the rheological behavior of the
widely studied steady-state flow, where the total two-phase flow rate varies
non-linearly with the excess pressure drop at low capillary numbers, and then
undergoes  a crossover to a linear Darcy regime at high Ca
\cite{tkr09,tlk09,aet14, rcs11,rcs14,sbd17,glb20, zbg21,zbb22}. The origin of
that non-linearity and the linear crossover has been well explained in terms of
the pore-scale disorders in the capillary pressures \cite{sh12,rhs19}, radii
distribution \cite{rsh21} and wettabilities \cite{fsr21,fsh23}. Though the
values of the steady-state non-linear exponent has been reported to differ for
different systems, the range of them was similar to what we find here in Fig.
\ref{fig_beta}.

\section{Discussion and conclusions} 
\label{sec_summary}
We have addressed in this article how the growth rate of fingers during
immiscible two-phase flow (drainage) in porous media depends on the local
viscous pressure drops between the two fluids. In other words, we revisited the
comparison of immiscible viscous fingers with Saffman-Taylor viscous fingering
and DLA, for which a linear Laplacian growth is assumed. We considered pore
networks with a regular geometry but with disorder in the pore/channel widths,
and focused on the intermediate to higher capillary numbers, which include the
intermediate regime between capillary and viscous fingers and the regime of pure
viscous fingering. First, we looked into the geometrical properties of the
drainage fingers, and compared their statistical shape in the reference frame of
the advancing finger to the smooth Saffman-Taylor fingers that develop in the
continuum scale description of two-phase flow in porous media. Saffman and
Taylor mentioned that viscous fingering in an empty Hele-Shaw cell without any
porous structure is equivalent to what should be expected in a porous medium
\cite{st58}. Though this is true for single phase flow, it is not necessarily
true for two-phase flow, except perhaps if the porous medium is a regular
lattice (no disorder), because the continuum description cannot account for the
role of surface tension in a disordered porous medium. Indeed, in an empty
Hele-Shaw cell, the fluid-fluid interface is continuous, and the long-range
in-plane component of its curvature induces long-range forces, whereas in a
porous medium the interface is broken up in many small menisci, and surface
tension acts at the scale of these menisci, which is the pore scale. The
in-plane component of the curvature is then much larger, and, furthermore,
stochasticity in the pore sizes plays an important role in shaping the
displacement process by inducing a stochasticity in the capillary pressure
thresholds.

We thus measured the statistical longitudinal volumetric density and growth rate
profiles of the fractal drainage fingers in the dynamic reference frame attached
to the most advanced finger tip, and compared them with the properties of smooth
Saffman-Taylor fingers and of DLA growth structures. The width ratio with
respect to the medium's width,  $\lambda$, of the fingers' statistical density
map, was found to approach the $0.5$ Saffman-Taylor value at intermediate
capillary numbers Ca, and then increase to around $0.8$ when either increasing
or decreasing the capillary numbers from that intermediate Ca range. The
corresponding longitudinal volumetric density and growth rate profiles behave
accordingly. For Saffman-Taylor fingers in continuum-medium, $\lambda$ was
observed to increase with the increase of surface tension \cite{cdh86,ms81,v83},
and it also increases with the angle of inlet wedge for systems with divergent
wedges \cite{aet96,acg89}. In case of our present system, it may be due to a
combined effect of these two, and needs further study.

The maximum growth happens at the most advanced fingertip, and decreases
exponentially behind it. Far behind the tip, the growth is almost frozen: the
interfaces between the two fluids are  held in place by the capillary barriers
and thus do not move. We show that this growth behavior is directly correlated
with the local pressure drop between the two phases, and controlled by the
distribution of the capillary forces. By computing numerically an effective
capillary threshold, we show that there exists a regime at intermediate
capillary numbers where the linear Laplacian growth property does not hold as
the local growth rate varies non-linearly with the excess local pressure drop
across the interface (i.e., the excess between the local porous pressure drop
and the capillary pressure threshold), $\Delta P$. This nonlinear regime of the
growth of the invading finger is explained by accounting for the disorder
distribution in the capillary thresholds in a theoretical assessment of the link
between $\Delta P$ and the local growth rate. Indeed, when $\Delta P$ falls
within the range of the capillary pressure thresholds, increasing the capillary
number means that not only will the invasion velocity increase linearly with
$\Delta P$, but the number of pores along the front that are invaded at any time
will also increase, thus rendering the growth rate nonlinear. For a uniform
capillary threshold distribution, the resulting growth law is expected to be
quadratic. In our disorder pore networks, that distribution is not uniform, and
consequently the nonlinear growth exponent $\beta$ depends on the capillary
number. The numerical simulations provide a range of $\beta$ values similar to
that of the rheology exponent previously reported for steady-state flow. In
fact, it transitions from a plateau at values as large as 2.5--3 at the smallest
investigated capillary numbers, to another plateau at 1 for the largest
investigated capillary numbers. This large capillary number limit of $\beta=1$
is to be expected, since it correspond to configurations in which $\Delta P$ is
larger than the upper boundary of the medium's capillary pressure thresholds; in
such configurations our theoretical assessment predicts a crossover of the flow
regime to the linear Laplacian growth.

This study opens many prospects for future studies. One of them is the
systematic investigation of the link between the capillary threshold
distribution of the porous medium and the nonlinear growth exponent $\beta$.
Another one would be to study whether the values of $\beta$ are comparable for
steady state flow and the present viscous fingering process in the same porous
geometry. In order for the distribution of the capillary barriers to only depend
on the pore network geometry, the fluid-fluid interface must sample the entire
barrier distribution at any time, otherwise the capillary barrier/capillary
threshold distribution will also depend on the flow patterns, in a manner that
may differ between the steady-state flow and viscous fingering configurations.

\linespread{1.22}
\section*{Acknowledgements}
This work is supported by the Research Council of Norway through its Centers of
Excellence funding scheme, project number 262644.

\section*{References}
%

\end{document}